# Pathologist-like explainable AI for interpretable Gleason grading in prostate cancer

Gesa Mittmann[1,2*], Sara Laiouar-Pedari[1*], Hendrik A. Mehrtens[1*], Sarah Haggenmüller[1], Tabea-Clara Bucher[1], Tirtha Chanda[1], Nadine T. Gaisa[3,4], Mathias Wagner[5], Gilbert Georg Klamminger[5], Tilman T. Rau[6], Christina Neppl[6], Eva Maria Compérat[7], Andreas Gocht[8], Monika Hämmerle[9], Niels J. Rupp[10,64], Jula Westhoff[11], Irene Krücken[12], Maximillian Seidl[6], Christian M. Schürch[14,15], Marcus Bauer[9], Wiebke Solass[16], Yu Chun Tam[17], Florian Weber[18], Rainer Grobholz[19,20], Jaroslaw Augustyniak[21], Thomas Kalinski[22], Christian Hörner[23], Kirsten D. Mertz[24,25], Constanze Döring[26], Andreas Erbersdobler[27], Gabriele Deubler[28], Felix Bremmer[29], Ulrich Sommer[30], Michael Brodhun[31], Jon Griffin[32], Maria Sarah L. Lenon[33,34], Kiril Trpkov[35], Liang Cheng[36], Fei Chen[37], Angelique Levi[38], Guoping Cai[38], Tri Q. Nguyen[39], Ali Amin[40], Alessia Cimadamore[41], Ahmed Shabaik[42], Varsha Manucha[43], Nazeel Ahmad[44], Nidia Messias[45], Francesca Sanguedolce[46], Diana Taheri[47,48], Ezra Baraban[49], Liwei Jia[50], Rajal B. Shah[50], Farshid Siadat[35], Nicole Swarbrick[51, 52], Kyung Park[53], Oudai Hassan[54], Siamak Sakhaie[55], Michelle R. Downes[56], Hiroshi Miyamoto[57], Sean R. Williamson[58], Tim Holland-Letz[59], Carolin V. Schneider[60], Jakob Nikolas Kather[61], Yuri Tolkach[62,63], Titus J. Brinker[1†]

## Affiliations

1. Division of Digital Prevention, Diagnostics and Therapy Guidance, German Cancer Research Center (DKFZ), Heidelberg, Germany
2. Medical Faculty of University Heidelberg, Heidelberg, Germany
3. Institute of Pathology, RWTH Aachen University, Aachen, Germany
4. Institute of Pathology, University Hospital, University of Ulm, Germany
5. Department of Pathology, University of Saarland, Homburg Saar Campus, Homburg Saar, Germany
6. Institute of Pathology, University Hospital Düsseldorf, Düsseldorf, Germany
7. Department of Pathology, Medical University of Vienna, Vienna, Austria
8. Institute of Pathology, University Hospital Schleswig-Holstein, Lübeck, Germany
9. Institute of Pathology, Martin Luther University Halle-Wittenberg, Halle (Saale), Germany
10. Department of Pathology and Molecular Pathology, University Hospital Zurich, Zurich, Switzerland
11. Institute of Pathology, Städtisches Klinikum Karlsruhe, Karlsruhe, Germany.
12. Institute of Pathology, University of Leipzig, Leipzig, Germany
13. Department of Pathology, University Medical Center Freiburg, Freiburg, Germany
14. Department of Pathology and Neuropathology, University Hospital and Comprehensive Cancer Center Tübingen, Tübingen, Germany

15. Cluster of Excellence iFIT (EXC 2180) "Image-Guided and Functionally Instructed Tumor Therapies", University of Tübingen, Germany
16. Institute of Pathology, University Bern, Bern, Switzerland
17. Institute of Pathology, Georgius Agricola Foundation Ruhr, Ruhr University Bochum, Bochum, Germany
18. Institute of Pathology, University of Regensburg, Regensburg, Germany
19. Institute of Pathology, Cantonal Hospital Aarau, Aarau, Switzerland
20. Medical Faculty University of Zurich, Zurich, Switzerland
21. Institute of Pathology, Kantonsspital St Gallen, St Gallen, Switzerland
22. Clinic for Pathology, University Hospital Brandenburg an der Havel, Brandenburg an der Havel, Germany
23. Institute for Pathology, University Medical Faculty Mannheim, University of Heidelberg, Manheim, Germany
24. Institute of Medical Genetics and Pathology, University Hospital Basel, Basel, Switzerland
25. Department of Biomedicine (DBM), Pathology of Infectious and Immunologic Diseases, University of Basel, Basel, Switzerland
26. Pathology Zwickau, Diagnosticum, Germany
27. Institute of Pathology, University Medicine Rostock, Rostock, Germany
28. Institute of Pathology, Kreiskliniken Reutlingen, Reutlingen, Germany
29. Institute of Pathology, University Medical Center Goettingen, Goettingen, Germany
30. Institute of Pathology, University Hospital of Dresden, Dresden, Germany
31. Institute of Pathology and Neuropathology, HELIOS Klinikum Erfurt, Erfurt, Germany
32. School of Medicine and Population Health, University of Sheffield, Sheffield UK
33. Department of Pathology, Keck School of Medicine, University of Southern California, Los Angeles, CA, USA
34. Department of Pathology and Laboratory Medicine, National Kidney and Transplant Institute, Quezon City, Metro Manila, Philippines
35. Department of Pathology and Laboratory Medicine, Cumming School of Medicine, University of Calgary, Rockyview General Hospital, Canada
36. Indiana University School of Medicine, Indianapolis, IN, USA
37. Department of Pathology, NYU Langone Health, New York, NY, USA
38. Department of Pathology, Yale University School of Medicine, New Haven, CT, USA
39. Department of Pathology, University Medical Centre Utrecht, Utrecht, The Netherlands
40. Department of Pathology, Warren Alpert Medical School of Brown University, Providence, RI, USA
41. Pathological Anatomy, University of Udine, Udine, Friuli-Venezia Giulia, Italy
42. Department of Pathology, UC San Diego School of Medicine, La Jolla, CA, USA
43. Department of Pathology, University of Mississippi Medical Center, Jackson, MS, USA
44. James A. Haley Veterans Hospital, Tampa, FL, USA
45. Department of Pathology and Immunology, Washington University in St. Louis, St. Louis, MO, USA
46. Department of Pathology, University of Foggia, Foggia, Italy
47. Department of Pathology, Kidney Diseases Research Center, Isfahan University of Medical Sciences, Isfahan, Iran
48. Urology Research Center, Tehran University of Medical Sciences, Tehran, Iran

49. Department of Pathology, Johns Hopkins University, Baltimore, MD, USA
50. Department of Pathology, University of Texas Southwestern Medical Center, Dallas, TX, USA
51. Department of Anatomical Pathology, PathWest Laboratory Medicine WA, Western Australia, Australia
52. Division of Pathology and Laboratory Medicine, UWA Medical School, University of Western Australia, Crawley, Western Australia, Australia
53. Department of Pathology, New York University Health School of Medicine, NY, USA
54. Department of Pathology, Henry Ford Health System, Detroit, MI, USA
55. Department of Anatomical Pathology, Frankston Laboratory, Dorevitch Pathology, VIC, Australia
56. Anatomic Pathology, Precision Diagnostics & Therapeutics Program, Sunnybrook Health Sciences Centre, Toronto, ON, Canada
57. Department of Pathology and Laboratory Medicine, University of Rochester Medical Center, Rochester, NY, USA
58. Robert J Tomsich Pathology and Laboratory Medicine Institute, Cleveland Clinic, Cleveland, OH, USA
59. Division of Biostatistics, German Cancer Research Center, Heidelberg Germany
60. Department of Internal Medicine, University Hospital Aachen, RWTH University of Aachen, Aachen, Germany
61. Else Kroener Fresenius Center for Digital Health, Faculty of Medicine and University Hospital Carl Gustav Carus, TUD Dresden University of Technology, Dresden, Germany
62. Institute of Pathology, University Hospital Cologne, Cologne, Germany
63. Medical Faculty, University of Cologne, Cologne, Germany
64. Faculty of Medicine, University of Zurich, Zurich, Switzerland

\* Contributed equally

**† Correspondence to:**

Titus J. Brinker, MD

Department of Digital Prevention, Diagnostics and Therapy Guidance, German Cancer Research Center (DKFZ), Im Neuenheimer Feld 223, 69120 Heidelberg, Germany
Tel.: +496221 3219304; E-Mail: titus.brinker@dkfz.de

# Abstract

The aggressiveness of prostate cancer, the most common cancer in men worldwide, is primarily assessed based on histopathological data using the Gleason scoring system. While artificial intelligence (AI) has shown promise in accurately predicting Gleason scores, these predictions often lack inherent explainability, potentially leading to distrust in human-machine interactions. To address this issue, we introduce a novel dataset of 1,015 tissue microarray core images, annotated by an international group of 54 pathologists. The annotations provide detailed localized pattern descriptions for Gleason grading in line with international guidelines. Utilizing this dataset, we develop an inherently explainable AI system based on a U-Net architecture that provides predictions leveraging pathologists' terminology. This approach circumvents post-hoc explainability methods while maintaining or exceeding the performance of methods trained directly for Gleason pattern segmentation (Dice score: $0.713_{\pm 0.003}$ trained on explanations vs. $0.691_{\pm 0.010}$ trained on Gleason patterns). By employing soft labels during training, we capture the intrinsic uncertainty in the data, yielding strong results in Gleason pattern segmentation even in the context of high interobserver variability. With the release of this dataset, we aim to encourage further research into segmentation in medical tasks with high levels of subjectivity and to advance the understanding of pathologists' reasoning processes.

# Introduction

Prostate cancer is a major health issue, affecting approximately 5 million men globally, with around 1.5 million new cases reported in 2020 [1]. The Gleason grading system, developed by Donald Gleason in 1974 [2] and most recently discussed and updated by the International Society of Urology Pathology (ISUP) and by the Genitourinary Pathology Society (GUPS) in 2019 [3,4], remains the primary method for assessing tumor aggressiveness and prognosis in patients with prostate cancer [25].

For Gleason scoring in the context of primary diagnosis, pathologists assess histological architectural features such as gland shape and size based on tumor biopsy samples and assign Gleason patterns ranging from 1 (resembling gland-like structures) to 5 (resembling least gland-like structures). Gleason patterns 1 and 2 were merged with pattern 3 in later modifications of the system [6], therefore the Gleason score, quantified as a sum of the most predominant and the highest Gleason pattern, ranges from 6 (3+3) to 10 (5+5). Higher scores indicate more aggressive tumors [7]. Despite its widespread use, however, the Gleason system has limitations, including sampling bias and subjective assessment of tumor architecture resulting in significant interobserver variability [8].

Multiple studies have shown that artificial intelligence (AI)-based image analysis has the potential to assist pathologists in Gleason grading, potentially matching or exceeding human accuracy [9–12]. Developing robust AI models for this task requires large datasets with expert annotations. For Gleason grading, datasets such as the Gleason19 Challenge [13,14] and the PANDA Challenge [11,12] are openly available; however, in most cases the available annotations indicate the area of patterns relevant to the final scoring or merely the Gleason score, without providing an explanation of the specific histological criteria behind the decisions. Consequently, the typical approach to Gleason grading with AI involves end-to-end models that predict Gleason patterns or even the score directly from the images. Although these models can achieve high accuracy, their decision-making process lacks transparency, which may present a barrier to clinical adoption [15,16], particularly in light of patients' right to explanation [17]. Especially in fields such as Gleason grading, where there is a significant subjectivity in the assessment [8], the interest in AI-assisted diagnostic systems is high. However, there is a demand for clear and reliable explanations [18].

In order to overcome interpretability issues for neural networks, post-hoc explainability techniques such as CAM or Grad-CAM [19], LRP [20], and LIME [21,22,23] have been developed that highlight regions of interest relevant to the decision made. These heatmaps aim to provide visual explanations for AI decisions, for example by indicating pixels that have a high influence on the predicted outcome. However, these methods often provide pseudo-explainability with vague morphological correlates. Interpreting the results requires specialized expertise [24], and it has been shown that using these approaches carries a high risk of confirmation bias [25,26], a cognitive tendency whereby individuals favor evidence that confirms their pre-existing hypotheses and beliefs. Additionally, such indicated regions may not always correspond to the actual causative regions of the cancer patterns [27,28], but might instead show unwanted statistical correlations learned by the neural network - a crucial factor that is rarely addressed. As pathologists prefer simple, visual explanations that are grounded in morphology and reflect their way of thinking, an inherently explainable approach to AI with clear and intuitive explanations is needed [29].

To address these significant limitations of traditional algorithm development, we propose the use of a concept-bottleneck-like [30] U-Net [31] architecture to develop a pathologist-like, inherently explainable AI system (GleasonXAI), as presented in **Figure 1**. For the development we compile and open source one of the largest datasets of annotations localizing explanations for Gleason patterns in tissue microarray (TMA) core images. The resulting GleasonXAI offers interpretability for AI-assisted segmentation of Gleason pattern by directly recognizing and delineating pre-defined histological features, which are associated with a textual explanation using terminology common to pathologists and rooted in GUPS and ISUP recommendations.

Utilizing novel approaches to train and evaluate models with soft labels, we capture the intrinsic uncertainty in the training data, thereby providing promising Gleason pattern segmentation in spite of high interobserver variability, and exceeding the performance of traditional approaches directly trained to predict Gleason patterns.

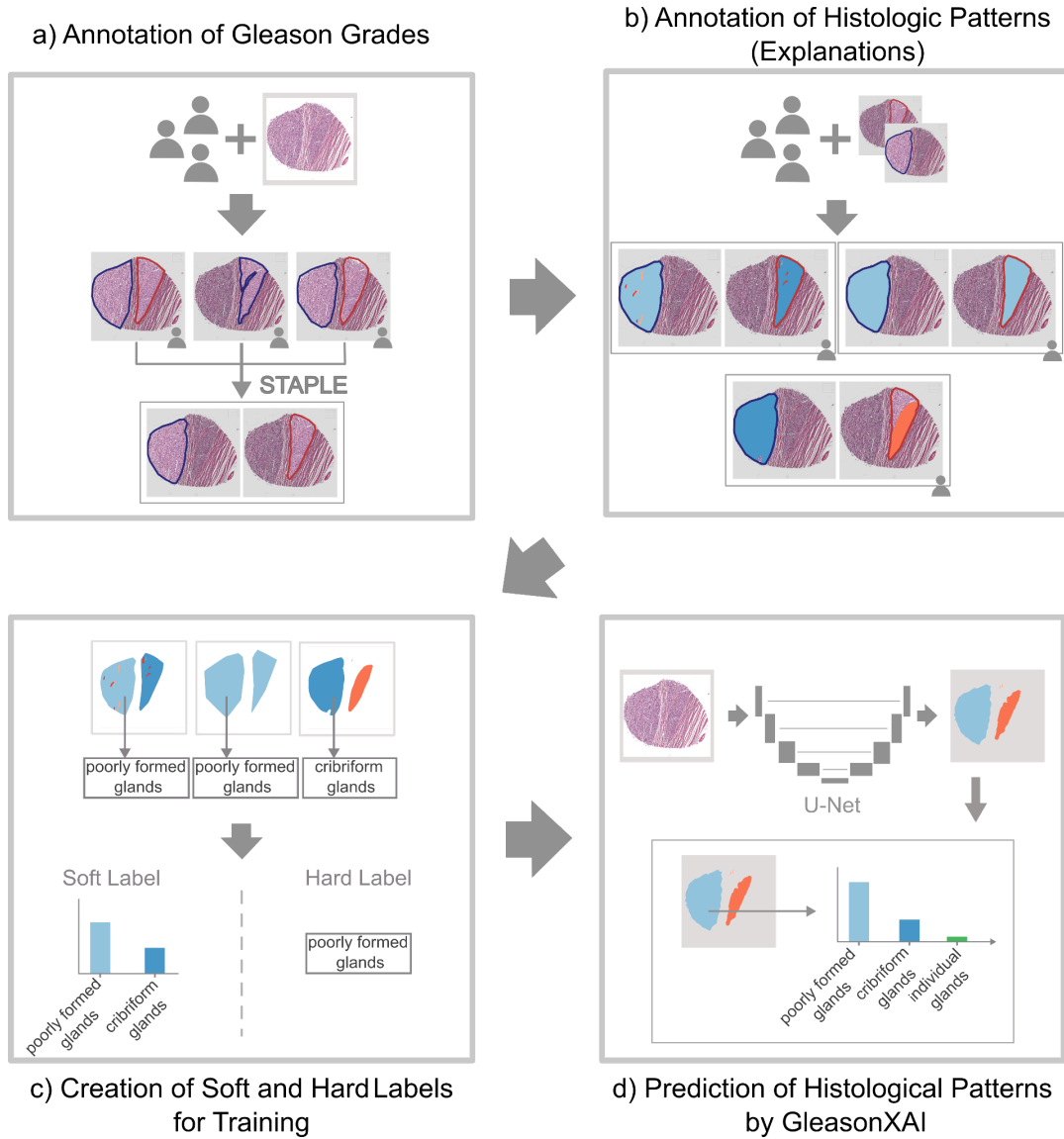

**Figure 1**. a) Up to six pathologists evaluated the TMA core images, identifying areas for each Gleason pattern, which were then merged using the STAPLE algorithm. b) Subsequently, three pathologists independently annotated histologic patterns based on a predefined ontology. c) We compared training on two labeling approaches: soft and hard labels. In the soft label approach, each pixel is represented as a distribution across the annotated classes, while the hard label method assigns a class to each pixel through majority voting. d) As a result, we developed GleasonXAI, a U-Net model that generates segmentation masks closely aligned with the pathologists' consensus. Due to training with soft labels, the predicted distributions also often reflect their agreement. Further details on post- and pre-processing such as the masking of background pixels can be found in the **Methods** section.

# Results

## Pathologist Characteristics

Between March 2023 and October 2023, an international team of 54 pathologists from ten countries participated in the study, with the majority of participants from Germany (22) and the USA (18). Among the participants, 47 were responsible for explanatory annotations, six for Gleason grade annotations, and one for creating the initial terminology, which was later reviewed and adapted in a panel of nine of the participants (see **[Methods](#)** section). The pathologists had a median of 15 years of clinical experience in pathology, with individual experience ranging from one to 35 years. Notably, 28 of these 54 annotators had extensive experience, defined as 15 years or more. In their clinical practice, the participating pathologists signed out a median of 15 prostate cancer cases per week, with individual prostate cancer caseloads ranging from fewer than ten to 75 patient cases weekly.

## Dataset Characteristics

The annotated dataset generated in this study comprised 1,015 TMA core images. The images were sourced from three distinct datasets, each created by a different institution. The annotations consisted of areas to which explanations describing histological patterns were assigned. Explanations could be mapped to one of the three Gleason patterns or further divided into more detailed sub-explanations, which described more specific subgroups of the histological features defined by their parent explanation. For additional information, please refer to the **[Methods](#)** section.

For each of the Gleason patterns, there were a considerable number of images containing associated annotations. Specifically, 55.76% of the images contained annotations for Gleason pattern 3 (566/1015), 74.48% for Gleason pattern 4 (756/1015), and 32.32% for Gleason pattern 5 (328/1015) (see **Figure 2a**). When analyzing the segmentation masks on both the explanation and sub-explanation level, the number of images containing the classes at least once exhibited a higher variation, with values ranging from 57 to 729 for explanations and 0 to 526 for sub-explanations (see **Figure 2b** and **Figure 2c**). However, the broader, classical explanations yielded a more balanced distribution, with fewer small classes.

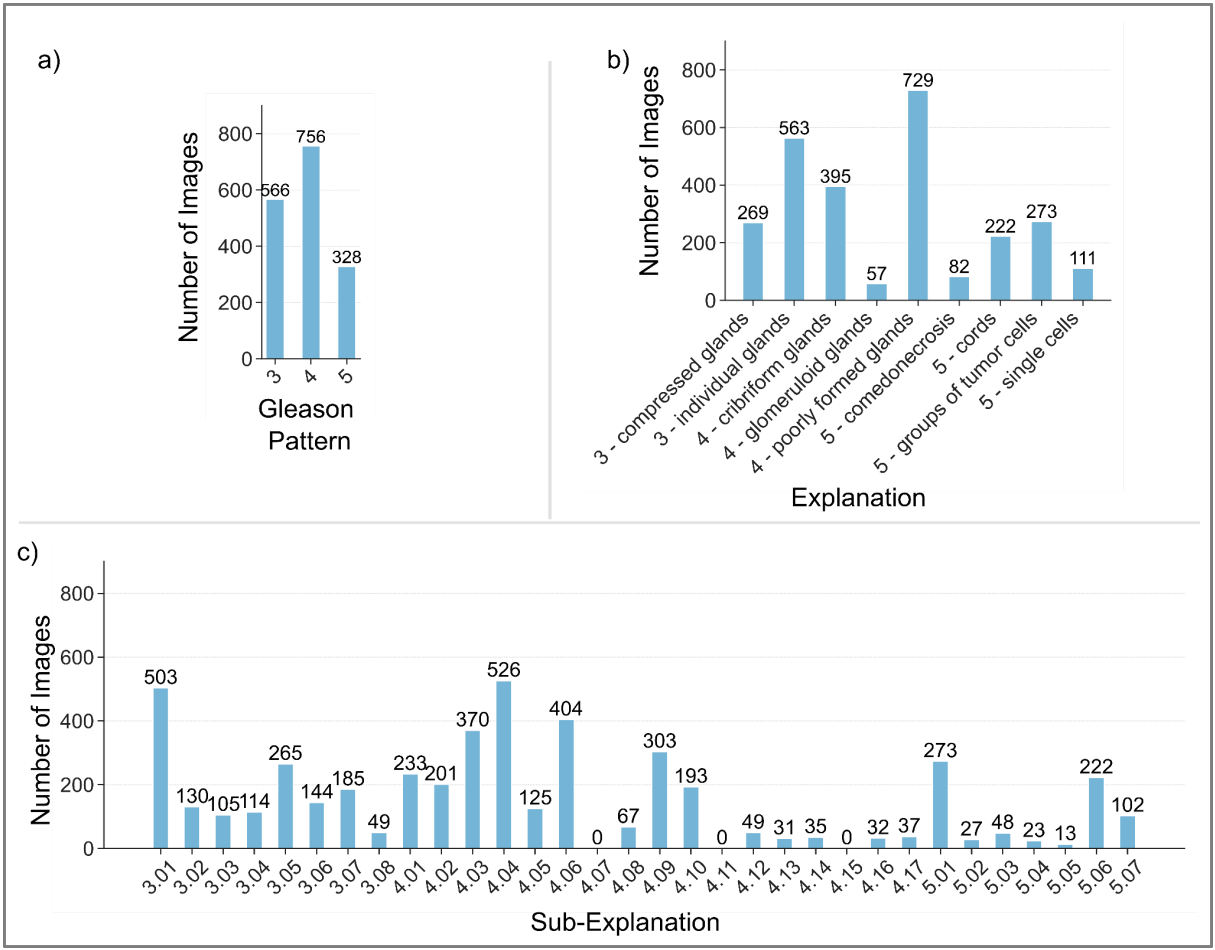

**Figure 2.** Class distribution. Number of TMA core images with at least one occurrence of a) the specified Gleason pattern, b) the specified explanation, and c) the specified sub-explanation. *Benign tissue* is not included, as it is present in all images. A mapping of sub-explanation numbers to text and is available in **Supplementary Table S.1**. The mapping of the explanations to their long version is available in **Figure 8**.

## Agreement between Pathologists Varies Depending on Histopathologic Pattern

Understanding the interobserver variability in the ground truth for Gleason grading was crucial for improving the reliability of our classifier, as inconsistent annotations can significantly impact model performance.

In our dataset, the images were accompanied by Gleason score information (see **Methods** section), generated for each core by a consensus between one to six pathologists, to provide guidance to the annotators. Pathologists were, however, encouraged to use explanations of different Gleason patterns, if they disagreed. Comparing the given grade information with the annotated Gleason patterns (see **Figure 3a),** the annotations largely aligned with the given grade. The majority of discrepancies occurred at the boundaries between Gleason patterns 3 and 4, and Gleason pattern 4 and 5 - an expected observation, as borderline cases are a known source of interobserver variability. The high agreement is also reflected in the Fleiss' kappa [32] values, which ranged from 0.23 to 1.00 within the annotator groups when identifying Gleason patterns (see **Figure 3b, top**), indicating *fair* to *perfect agreement* according to Landis and Koch [33].

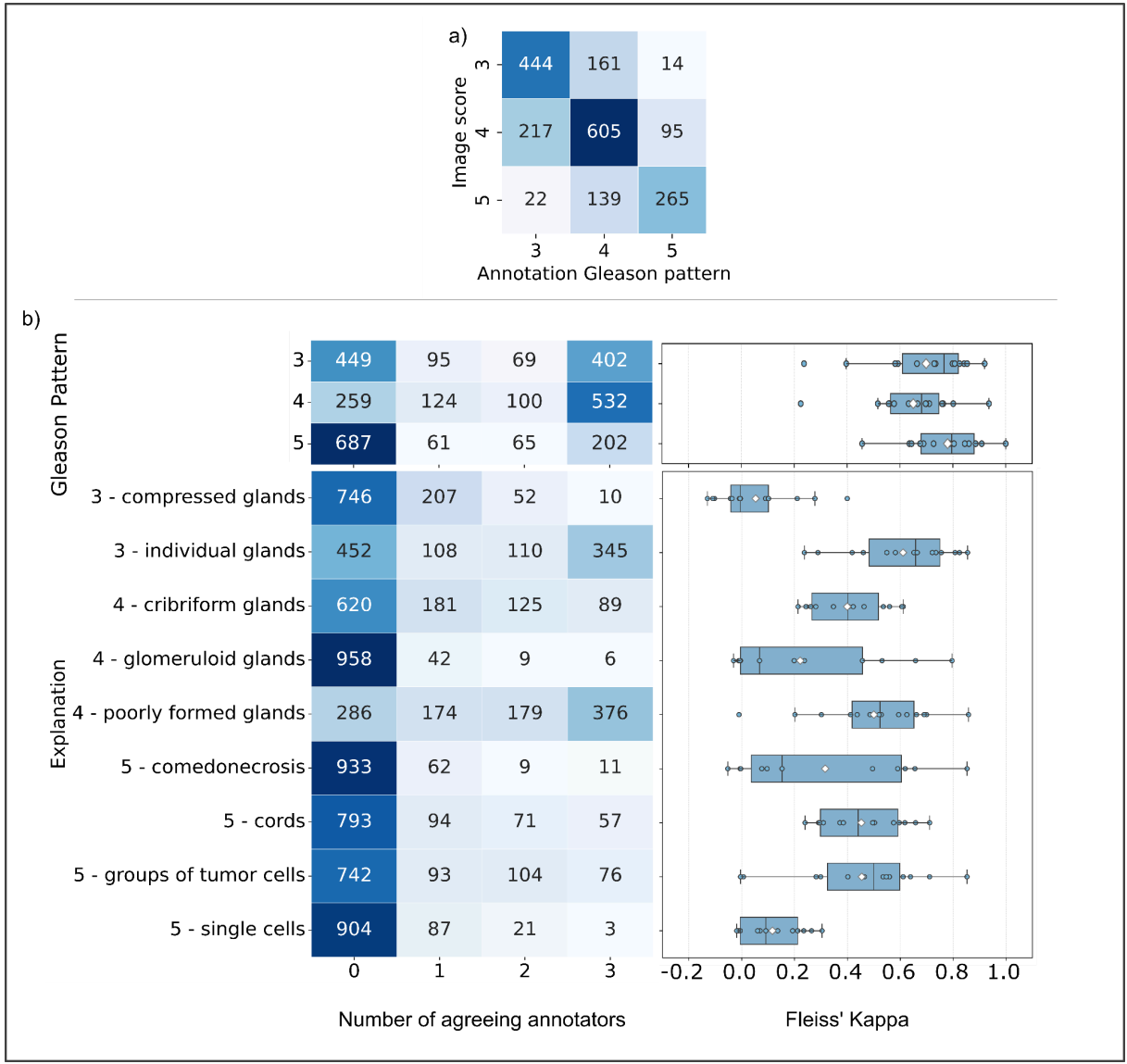

**Figure 3.** Agreement of annotators for explanations on the image-level. a) Confusion matrix between given Gleason score and Gleason pattern of the applied explanations in the images (Gleason pattern was counted only once per image, regardless of number of agreeing pathologists), and b) heatmap containing the number of TMA core images in which n out of the three annotators indicated the presence of the Gleason pattern (left, top) explanations (left, bottom) and the resulting Fleiss' kappa for groups of three raters (on the right). Dots represent the groups of annotators, boxes represent quartiles, white diamonds the mean value, and whiskers extend to the furthest datapoint within 1.5 of the inter-quartile range. As not all groups used all categories, the number of groups per category varies. Precise numerical values can be found in **Supplementary Table S.5** to **S.7**, and for explanations in **Supplementary Figure S.1**. The mapping of the explanations to their long version is available in **Figure 8**.

Consensus on the specific histological patterns, however, and consequently the appropriate sub-explanations and explanations, was less frequent. The extent of this variability differs depending on the specific explanation under consideration.

An analysis of the distribution of the images across the number of annotators that agreed on the presence of each explanation (see **Figure 3b, left**) revealed high levels of agreement for certain histological features, such as *poorly formed glands* and *individual glands*. Specifically, 76.13% (555/729) of the images annotated with poorly formed glands and 80.81% (445/563) with individual glands by at least one annotator reached at least two-rater agreement. This is further supported by their respective mean Fleiss' kappa values of $0.50_{\pm 0.23}$ and $0.61_{\pm 020}$ (see **Figure 3b, right),** indicating moderate to substantial agreement.

Conversely, there are also explanations, such as *glomeruloid glands* and *single cells*, where it was rare for a second or third annotator to agree (see **Figure 3b, left**). Subgroup analyses identified particularly pronounced interobserver variability for the explanations of *single cells and compressed glands,* with Fleiss' kappa values of 0.145 and 0.180, respectively (see **Supplementary Table S.2**). The explanations of *glomeruloid glands* and *comedonecrosis* on the other hand exhibited large variance in Fleiss' kappa values across the annotator groups, ranging from -0.031 to 0.796 and -0.052 to 0.852, respectively. Notably, with the exception of *compressed glands*, these explanations were the rarest annotated classes (see **Figure 2).**

The agreement on the sub-explanation was notably low, with Fleiss' kappa values ranging from -0.22 to 0.85 between the groups, as illustrated by the predominantly *slight agreement* shown in **Supplementary Figure S.1** for each label. As interobserver agreements for most histologic patterns, which are equivalent to our explanations, are reported to be *fair* or *moderate* [34–36], it is to be expected that agreement on even finer details of the patterns will be lower. These results indicated considerable noise in the identification of sub-explanations, confirming the necessity of using the explanations that consolidate the detailed sub-explanations into broader, medically coherent categories. This step was crucial to reduce variability and to ensure more reliable training of our classifier.

Further details on the Fleiss' kappa values and their bootstrapped confidence intervals can be found in **Supplementary Table S.2** to **S.7**.

## Pixelwise Agreement Between Raters Is Lower in Minority Classes

Since the AI was tasked with learning the localization of the explanations, the annotators' agreement at pixel level was crucial. As their annotations served as the predictive targets for the AI, the level of interobserver agreement induced an upper limit on the performance the AI could reach.

A similar pattern of decreasing annotator agreement with increasing explanation detail was observed when analyzing the number of pixels with a unique majority vote. Of the 58.12% of pixels constituting the foreground of the dataset, 97.54% could be assigned a unique majority class when evaluating the Gleason patterns (36.23% with two rater and 61.30% with three rater agreement). However, at the explanation level, this dropped to 86.41% (41.07% with two rater and 45.35% with three rater agreement, respectively), and further to 67.76% at the sub-explanations level (37.80% and 29.96%, respectively), indicating a considerable proportion of pixels with a high annotation uncertainty.

It is worth noting that the classes with the lowest number of annotated pixels (see **Figure 4b**) were also precisely the classes where annotators demonstrated the least consensus. As illustrated in **Figure 4a)**, this was particularly evident for the explanations of *comedonecrosis*, *single cells*, *glomeruloid glands*, and *compressed glands*, where 88.46% to 94.96% of all annotated pixels were annotated by a single rater.

The lower number of pixels with a unique majority vote indicated that the agreement at the pixel level was weaker compared to the image level. Consequently, it could be inferred that Fleiss' kappa at image-level represented an upper bound on the overall agreement.

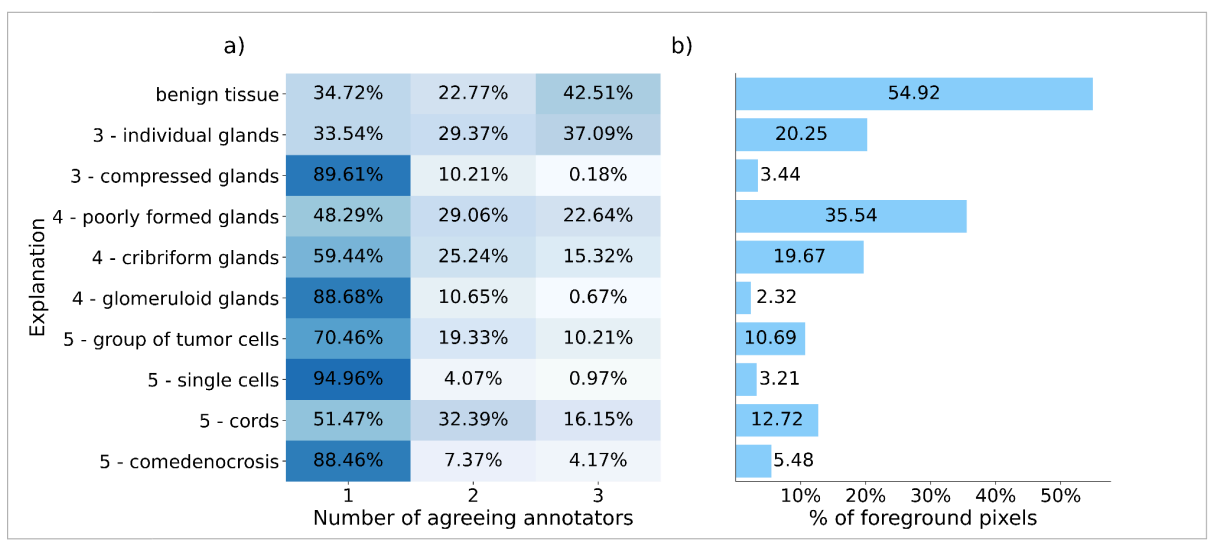

**Figure 4**. Agreement of annotators for explanations on pixel-level, demonstrated by a) the proportion of pixels of those assigned the explanation distributed according to the number of agreeing annotators, and b) the percentage of foreground pixels annotated with an explanation by at least one annotator. The mapping of the explanations to their long version is available in **Figure 8**.

## Model Development and Evaluation

### Soft Labels Improve Model Performance By Considering Annotator Uncertainty

To develop a pathologist-like, inherently explainable AI system for Gleason pattern segmentation (Gleason XAI), we selected a soft label approach, by treating the different annotations from different annotators over the pixels as probability distributions. This approach accounted for the high interobserver disagreement stemming from the reviewers' annotation uncertainty. Preserving all annotations instead of merging them with a traditionally used majority vote ensured that the AI system could also reflect the nuances of expert judgment. For comparison we also included classical hard label approaches, using the majority votes of our international expert team of pathologists.

For both approaches, we compared our models trained on the training data with different loss functions (see **Methods** section). Specifically, for the soft label approach, we used the cross-entropy loss and our custom SoftDiceLoss, while for the hard label approach, we employed the original Dice loss and cross-entropy loss (see **Figure 5**). All approaches were compared using the Macro SoftDice, $L_1$-norm, Dice and Macro Dice metrics on a holdout test set. Due to the large interobserver disagreement and class imbalance present in the sub-explanations, we trained our models on the broader, medically coherent explanation level of our ontology (see **Methods** section, **Figure 8**). A comparison of all methods, when trained on the sub-explanations, as well as the full numerical results can be found in **Supplementary Figure 2** and **Supplementary Table S.8**.

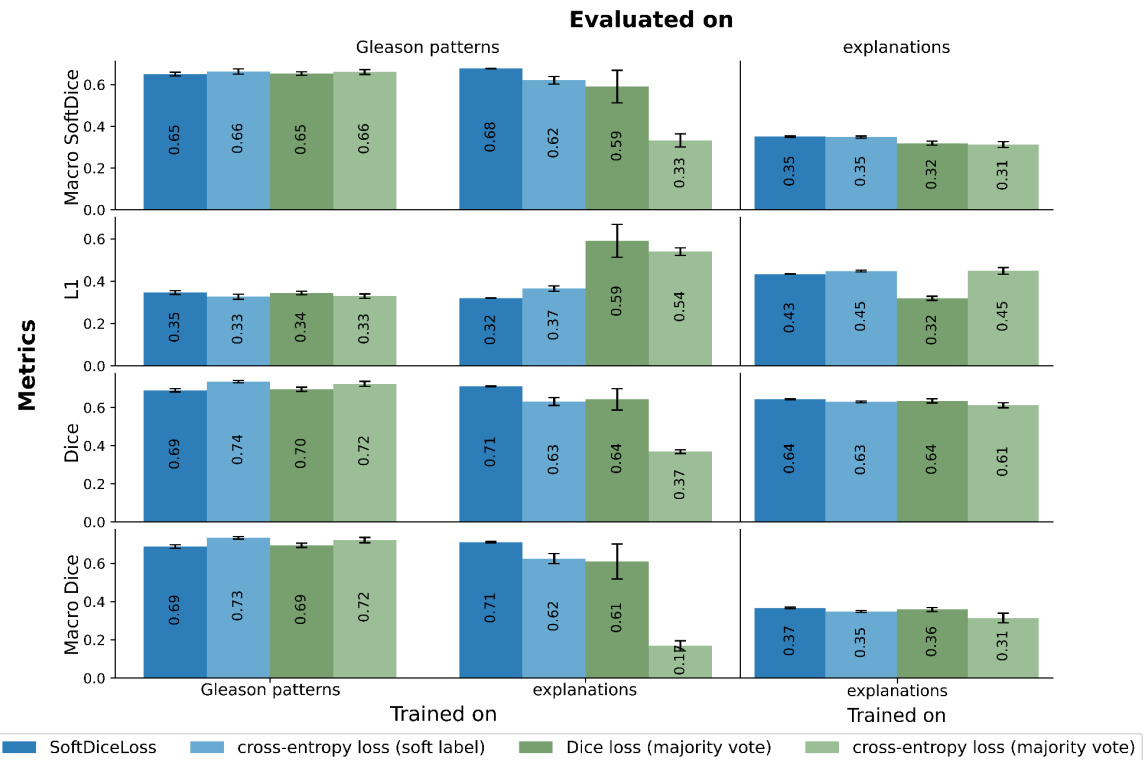

**Figure 5**. Results for our models trained with different loss functions, evaluated on the Gleason patterns and the corresponding explanations. Using our ontology, we mapped the labels upwards, allowing a comparison between the models trained on the explanations with those directly trained on the Gleason patterns. The bar plots display both the mean and the standard deviation, with the mean values additionally indicated within the bars. The green bar charts represent metrics for the hard label approaches, while the blue bars correspond to the soft label approaches. For the Dice metrics, higher values indicate better performance, while for the $L_1$-norm, lower values are preferable.

For most models, the metrics showed a decline from the evaluation on Gleason patterns to the explanations. This deterioration could be attributed to the increased number of classes, greater class imbalance in the segmentation task, and higher inter-rater variability, particularly affecting minority classes. Due to the high class imbalance in the explanations, this trend was especially noticeable in the class-balanced metrics (i.e Macro Dice and Macro SoftDice).

Training on majority-voting based explanation labels, as opposed to soft labels, resulted in a slight decrease in segmentation quality when evaluated on the explanations, especially when comparing the soft-label approaches with the cross-entropy loss. The difference became much more pronounced when the predictions on the explanations were mapped to the higher level Gleason patterns, where models trained on majority-voted labels performed worse than those trained using soft labels. Trained on the Gleason patterns, the hard and soft label approaches performed on par, with the cross entropy loss on the soft labels achieving the highest Dice and Macro Dice. Overall the soft label-based approaches consistently demonstrated superior performance in terms of the segmentation metrics on the test data compared to hard label approaches, when trained on the explanations.

Importantly, we were able to preserve the segmentation quality of the Gleason patterns even when training on the explanations. Models using the SoftDiceLoss, trained on the explanations but evaluated on the Gleason patterns, performed just as well as those trained directly on the Gleason patterns. This was reassuring, as it demonstrated that the inherent interpretability of our method did not come at the cost of reduced segmentation performance for the clinical task.

Similarly to the segmentation performance, models trained on explanations with our custom SoftDiceLoss exhibited better calibration to the pathologists' annotation distribution compared to the cross-entropy models trained with soft labels, as evidenced by a lower $L_1$-norm (see **Figure 5**).

Since the models trained with SoftDiceLoss on the explanations performed best across most metrics — except for the $L_1$-norm on the explanations — on both the explanations and Gleason patterns, we will use these models for the remainder of the analysis and define them as our GleasonXAI models. Further discussion on the calibration can be found in the **Supplementary Material Calibration Metric Discussion**.

## GleasonXAI Strongly Aligns With Pathologists

To further analyze the performance capabilities of our GleasonXAI models, we examined the distributions of predictions and the confusion matrices of the models (see **Figure 6**) on the test set. We combined the results of the three models by averaging the predictive distribution and confusion matrices.

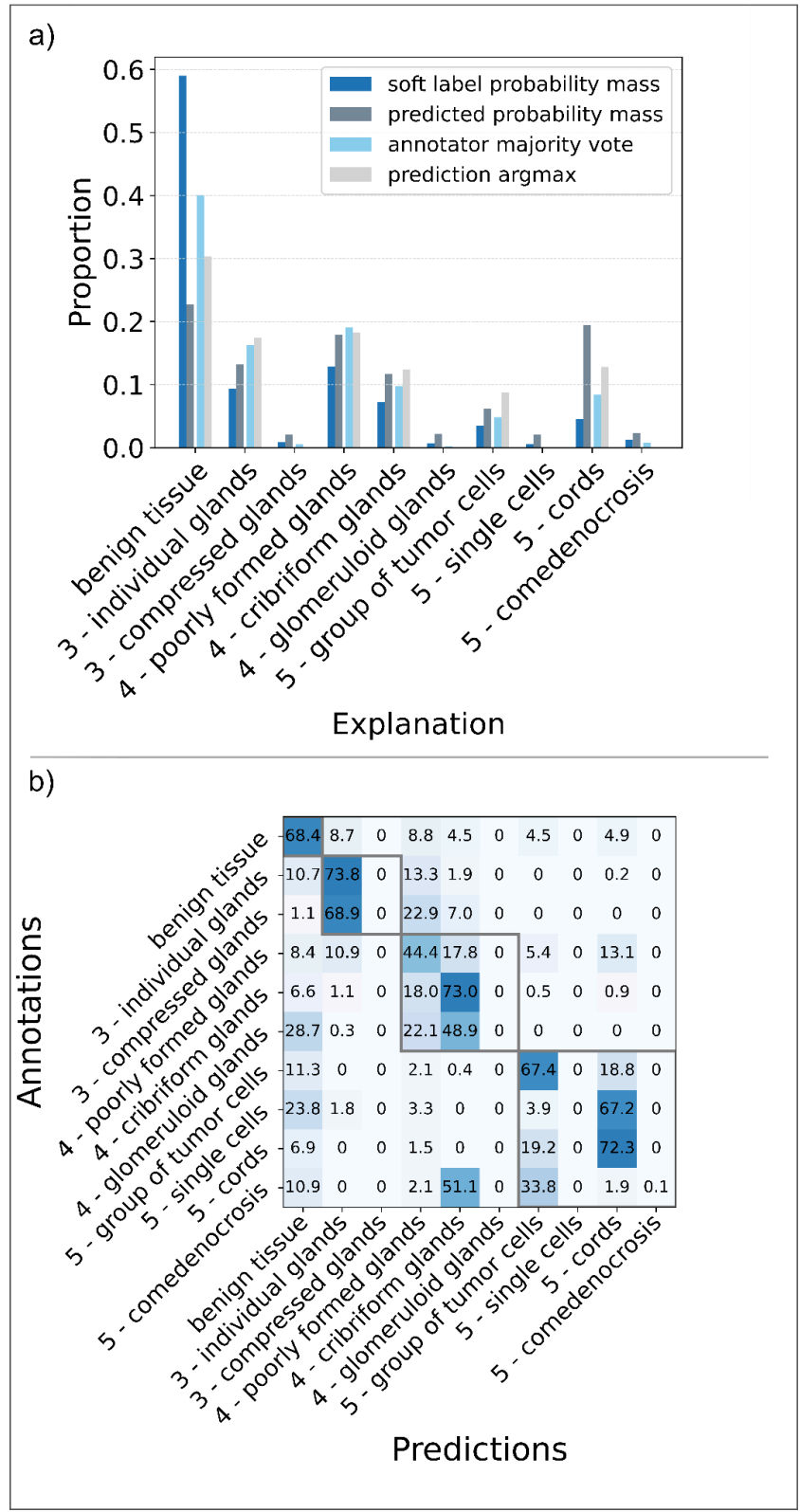

**Figure 6.** Comparison between GleasonXAI predictions and pathologist's annotations described by a) Proportion of predicted probability mass for each explanation, compared to the soft-label probability mass and the proportion of pixels with a majority vote for an explanation compared to the number of pixels with an argmax prediction for this explanation, and b) the confusion matrix for the argmax prediction and the majority label, presented in percentages. The gray boxes highlight the explanations corresponding to a common Gleason pattern. The mapping of the explanations to their long version is available in **Figure 8**.

Our GleasonXAI models predicted the majority of classes in the dataset with high reliability, often closely matching the prediction frequency of the annotations. Classes that were rarely annotated were predicted more frequently than they were annotated, and were also assigned a greater probability mass (see **Figure 6a**). However, for the rarest explanations — such as *glomeruloid glands*, *single cells*, and *presence of comedonecrosis* — the methods were not able to produce predictions with high certainty, likely due to their rarity in the training data. Similarly, the *comedonecrosis* class was rarely predicted. However, probability mass was assigned to these, which indicates that they are not fully unrecognized.

The confusion matrix in **Figure 6b** illustrates the annotations versus the predictions, revealing minimal confusion between explanations of different Gleason patterns, as indicated by the gray boxes, which underscored the strong performance of our models trained for Gleason pattern segmentation. Notably, explanations of Gleason patterns 3 and 5 were rarely misclassified as one another. Misclassifications predominantly occurred between adjacent Gleason patterns. This outcome was expected, as many cases could be medically categorized as falling between the Gleason pattern stages, and aligned with the pathologists' grading discrepancies shown in **Figure 3a**, where they also mostly differed between adjacent classes.

The class which was most frequently falsely classified was *benign tissue*, which we hypothesize is likely due to label uncertainty in the border regions of the annotations. This uncertainty might have arisen from the inherent difficulty in precisely determining annotation boundaries. Despite this, most of the predicted classes were predicted with high accuracy, with the lowest accuracy observed for the Gleason pattern 4 explanation of *poorly formed glands* at 44.4 % (± 1.18% SD) and the highest for the explanation of *individual glands* at 73.8 % (± 2.66% SD).

## GleasonXAI Generates Detailed Segmentation Maps

As segmentation maps that achieve high Dice scores can still exhibit unwanted properties like visual artifact or clutter [37], we qualitatively verified the correctness of our segmentation maps by visualizing them (see **Figure 7**). For these visualizations of the test data, we averaged the predictions of the three GleasonXAI models.

The models generally aligned well with the provided annotations, often producing segmentation maps that integrate elements from each individual annotation. This alignment was not restricted to the majority classes; the model also effectively captured fine details of less prevalent classes, even when there was no consensus among annotators on the presence of certain patterns. Notably, our model frequently generated more detailed segmentation masks than those in the reference testset (see **Figure 7g**), accurately predicting smaller details that were not annotated by multiple pathologists, confirming the high quality of our segmentations. These findings were encouraging as the model was able to also capture fine structural details and was not distracted by more coarse annotations.

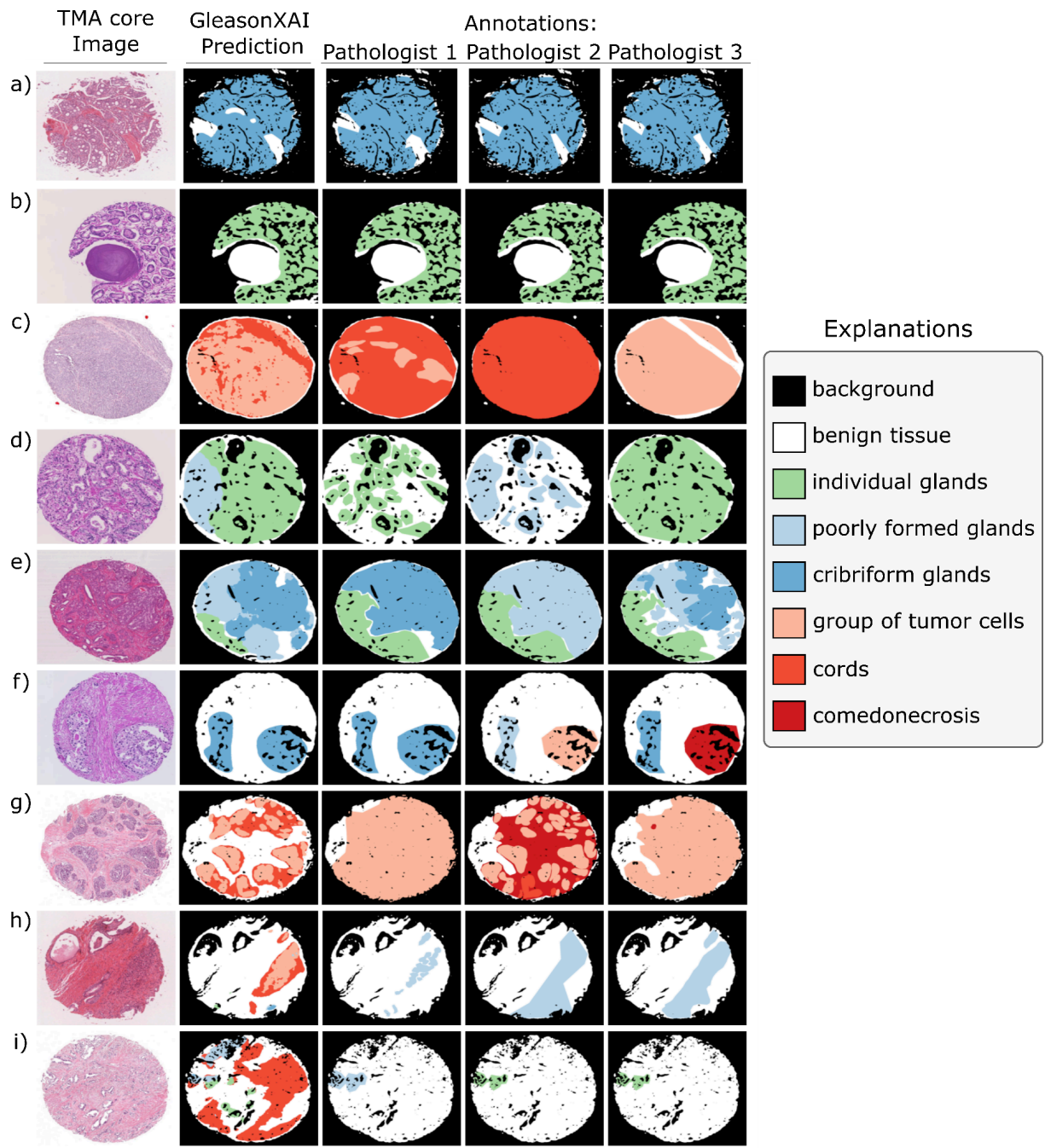

**Figure 7.** Visualization of segmentation results for the GleasonXAI model compared to pathologists' annotations. The segmentation images depict the argmax of the per-pixel distribution for the predictions of the model. Panels a-b) showcase examples of high agreement between the annotators and the model. Panels c-g) highlight cases with greater disagreement among the annotators, where the segmentation maps of the model often fell between the annotators' interpretations, reflecting the training objective of our soft-label approach. Panel h-i) illustrate instances of strong disagreement between the model and the annotators. Green labels belong to Gleason pattern 3, blue to Gleason pattern 4 and red to Gleason pattern 5. The mapping of the explanations to their long version is available in **Figure 8**.

# Discussion

The development of trustworthy and explainable machine learning models is crucial for their adoption in clinical practice [18]. However, previous studies have focused on detecting Gleason patterns using an end-to-end AI approach. This provides limited explainability, if any, through the use of post-hoc methods, which leaves room for potential interpretation and confirmation biases [28]. The goal of this study was to develop a pathologist-like, inherently explainable model for the segmentation of Gleason patterns, directly trained on annotations of morphological concepts collected through a collaborative effort of a large international group of 54 pathologists. In addition, we provide access to this large annotated dataset of TMA core images (n=1015) and detailed localized explanations for Gleason patterns.

By utilizing novel soft-label loss functions, such as our custom SoftDiceLoss, along with tailored assessment metrics, we were able to successfully train an AI system on a segmentation task characterized by substantial interobserver variability. The resulting AI is inherently explainable, directly providing explanations for tissue characteristics in accordance with the WHO/ISUP guidelines [6]. It demonstrates reliable performance across all but the rarest classes, which simultaneously exhibited highest discordance (e.g. *comedonecrosis, glomeruloid glands* and *single cells*), often annotated only by a minority of the raters.

Remarkably, these results were achieved without any loss in the segmentation quality for Gleason patterns compared to conventional methods trained directly on them (see **Figure 5**, $mean\,Dice_{\pm SD}$: $0.713_{\pm 0.003}$ vs. $0.691_{\pm 0.010}$). This is particularly encouraging, as inherently explainable XAI methods often suffer from an unfavorable performance-interpretability trade-off [38]. By treating the per-pixel annotations of multiple pathologists as soft-labels, rather than relying on high-variance, majority-voted labels, we were able to improve segmentation performance compared to hard label-based approaches, while also preserving and respecting the inherent uncertainty and ambiguity in Gleason pattern explanations.

We hypothesize that the benefits of using soft labels become more pronounced in scenarios with greater class imbalances and increased label uncertainty. This is especially visible when directly comparing the Macro Dice scores of cross entropy models trained on soft labels ($mean_{\pm SD}$: $0.625_{\pm 0.032}$) against those trained on majority-voted explanations labels ($mean_{\pm SD}$: $0.168_{\pm 0.033}$), when evaluated on the Gleason patterns. The considerable interobserver disagreement resulted in substantial sample variance for the majority-voted labels, contributing to label-noise, while additionally requiring the exclusion of 13.59% of all foreground pixels for the explanations (see **Figure 4**). By utilizing soft labels, we are able to incorporate all annotations — including minority opinions or classes that would otherwise be discarded in majority voting, or cases with multiple medically plausible annotations. This approach allows us to retain every pixel and provides an estimated annotation confidence, resulting in more conservative and distributed predictions and better predictions for minority classes.

Consequently, when training on Gleason patterns, the soft labels and the SoftDiceLoss did not provide a great advantage over the hard label-based approaches. Due to the relatively high level of agreement among pathologists on Gleason pattern level (see **Figure 3a**), many soft labels closely match with the majority-voted labels, and the class distribution is more balanced. This observation aligns with recent literature [39], which suggests that label smoothing — a label augmentation technique that produces distributional labels — is particularly beneficial in settings characterized by high label noise and class imbalance.

GleasonXAI was not able to predict the rarest of classes in a majority vote evaluation (see **Figure 6**), among them the morphological finding of *comedonecrosis*, which is an important histologic feature pathologists must evaluate during risk stratification [40]. We attribute this to their extreme rarity and the high inter-rater variability, even resulting in infrequent majority annotations among pathologists for these classes (see **Figure 3** and **Figure 4**). As a result, these classes were consistently ranked as the second most likely or lower in the predictions. However, aside for the rarest classes, the remaining minority classes were predicted with high accuracy, often more frequently and with greater probability mass than they were annotated (see **Figure 6a**). This behavior may be attributed to our class-averaged loss function, which emphasizes performance for minority classes or due to smaller, unannotated structures in the image, that were nonetheless predicted by GleasonXAI. Future studies could build upon this work by performing a targeted data collection for these less frequent explanations to further enhance the clinical utility of the model.

Whereas the pathologists achieved agreement similar to the literature in Gleason pattern annotation and some of the explanations [35,36,41], our analyses also revealed a higher-than-expected level of disagreement in others, such as single cells and comedonecrosis [34]. While each image was annotated by three expert pathologists — thereby exceeding the standard of care in terms of the number of observers — increasing the number of pathologists per image could therefore further improve the estimation of the underlying diagnostic distribution for each location. This would not only reduce sampling noise in the annotations, thereby improving the learning signal for hard label approaches, but also provide more precise estimates of diagnostic uncertainty. Such improvements would yield better and more continuous targets for the soft label approaches and allow for finer evaluation of the corresponding metrics.

Our work revealed a blind-spot in segmentation research using soft-labels. While recent segmentation loss-functions have been developed for training with soft labels [42], the evaluation and the presentation of results for pathologists still hinges on hard labels. Even with a perfectly estimated diagnostic distribution, a learned minority opinion within the diagnostic distribution would not be reflected in metrics based on hard labels or in visualizations that present only the most likely explanation of the predictive distribution. Since the goal of the study was to develop a model that closely matches the pathologists' consensus, addressing this challenge is beyond the scope for this paper. However, future work on the use of predictive distributions and soft labels in medical segmentation tasks is crucial. A potential approach could involve threshold based multi-label approaches or adapting conformal prediction techniques [43] to segmentation tasks.

The primary goal of our study was not to achieve state-of-the-art performance in Gleason pattern segmentation or grading, but rather to introduce a new approach for inherent and reliable explainability in Gleason pattern segmentation using novel annotations. Our approach could likely benefit from modern techniques like semi-supervised pre-training, even larger datasets, or more advanced training techniques like additional augmentations or ensembling. We encourage other researchers to further explore and improve upon this challenging segmentation task.

Recognizing the critical importance of explainable AI for the clinical adoption of automated Gleason grading, we developed GleasonXAI — a pathologist-like, inherently explainable model for the segmentation of Gleason patterns, trained directly on annotations created in collaboration with an international consortium of 54 pathologists. To further advance research on high-discordance medical datasets, we provide access to the largest annotated dataset of localized explanations for Gleason patterns, comprising 1,015 TMA core images.

# Methods

## Development of an Explanatory Ontology

To gather meaningful explanations for the dataset, we developed a comprehensive medical ontology with detailed explanations for the Gleason patterns 3, 4, and 5, based on the histological description of *Iczkowski KA*. for Gleason grading [44]. In collaboration with an experienced uro-pathologist (NTG), these explanations were shortened, split into distinct classes and translated into German for our German collaborators (see **Supplementary Table S.1**). Our main goal was to establish a criteria-based ontology for each Gleason pattern, incorporating distinct characteristics unique to each specific pattern, that should be later annotated by experienced pathologists.

Additionally, we conducted a panel discussion with expert uro-pathologists (n=9) to gather feedback on our approach. A key outcome was the need to adapt our ontology to current ISUP/World Health Organization (WHO) terminology. The wording changes were made while preserving the original content, resulting in our adapted ontology (see **Figure 8**).

For our model development we defined three distinct levels: Each Gleason pattern (depicted in dark blue) was assigned a set of broader, medical coherent explanations (depicted in light blue), which themselves contained multiple sub-explanations (depicted in white). These sub-explanations represent the original annotations that were gathered.

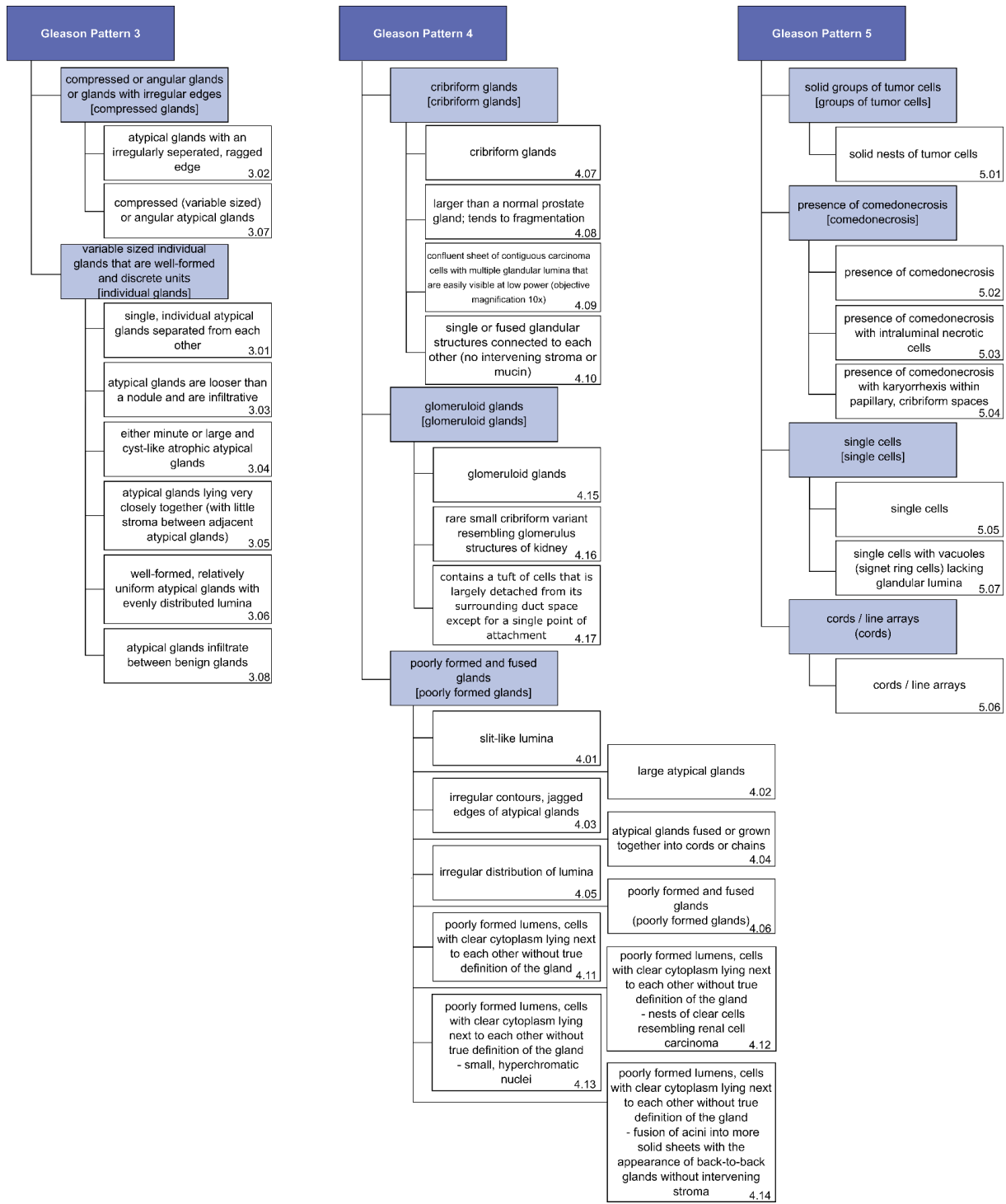

**Figure 8. Overview of the explanatory Gleason pattern ontology.** Generic terms based on WHO and ISUP2014 guidelines summarize the explanations corresponding to our initial ontology version. As the term "hypernephroid pattern" is now discouraged, we replaced it with "poorly formed lumens, cells with clear cytoplasm lying next to each other without true definition of the gland". Gleason pattern classes are marked in dark blue, explanation classes in light blue and sub-explanations in white. For our figures we use shortened names of the explanations which are shown in square brackets for explanations, and in the numbering in the bottom right for the sub-explanations.

## Utilized Datasets

We utilized data from three different data sources: 423 TMA core images were received from TissueArray.com [45], 538 from Arvaniti et al. Harvard Dataverse [46,47], and 219 from the Gleason19 Challenge [13,14]. The datasets were filtered to match our requirements of containing prostate adenocarcinoma tissue with Gleason Patterns 3, 4, and 5. In total, 1180 TMA cores of prostate adenocarcinomas were identified as eligible for annotation with detailed explanatory features.

## Annotating Procedure

To ensure high-quality annotations, we recruited an international team of 54 pathologists from university clinics, non-university public clinics, and private pathology practices through the ISUP platform or direct email invitations. Of the 54 annotators, 53 took part in the annotation process, which involved two tasks: First, three to four pathologists annotated the most prevalent and malignant Gleason pattern in the TMA core images; second, three annotators applied the explanatory ontology using the sub-explanations to generate detailed and explainable annotations.

The first task was only conducted on the TissueArray.com dataset, as the Gleason 19 challenge and Harvard Dataverse datasets already contained annotations of the Gleason patterns. The annotators were informed of the Gleason grading from the metadata of the Harvard Dataverse dataset (generated by a single pathologist based on hematoxylin and eosin staining and Immunohistochemical tests [45]), but could specify alternative patterns if they disagreed with the provided grading. The Gleason grade annotations from the pathologists were then merged using the STAPLE algorithm [48]. Similarly, the provided annotations for the Gleason19 Challenge dataset (generated by up to six pathologists [14]) were merged using the same algorithm. The resulting output masks were reviewed for quality and filtered by an observer (SLP).

After the merging of the Gleason Grade annotation masks, additional filtering was required. Images were removed (n=143) due to missing annotators, small or no Gleason grade areas, containing only Gleason Grade 1 or 2 post-merge, or other quality concerns. In 19 cases in the Gleason 19 Challenge dataset, the merged grade annotations produced by STAPLE did not align with any meaningful biological patterns (e.g too small or fragmented areas), though individual annotator labels did. For these cases, the annotation of the pathologist, whose annotation was deemed the closest match to the STAPLE output by an observer (SLP), was selected (see **Supplementary Figure S.3** for representative image).

For the second task, the TMA core images and their corresponding Gleason pattern annotation maps from the first task were divided in 15 distinct sets. Each set was provided to a group of annotators, who were then tasked with annotating specific histological patterns within the predefined areas to explain the respective Gleason pattern and assign a corresponding explanatory text. A free-text option was available if the pathologists disagreed with the provided explanation choices. For each annotated image from the first task, the pathologists of the second task received up to two different images, each with the outline of the annotation area for a single Gleason pattern (single grade images). In cases where the assigned Gleason grades were identical (i. e., 3+3, 4+4, or 5+5), only one single image with the corresponding area marked was presented (see **Supplementary Figure S.4**).

After the annotation with the explanatory labels, additional images (n=162) had to be excluded due to an insufficient number of raters. The objective was to obtain annotations from three pathologists for each image. We used the annotations of the first three pathologists who responded and completed their tasks. Pathologists who dropped out early in the annotation process were replaced, and their contributions were not included in the final dataset, if they annotated less than a quarter of their assigned data. The images for which we did not receive full annotations from three pathologists by the end of the process were removed. This occurred primarily due to late dropouts, but also due to single images being skipped. We reviewed the skipped images to account for potential systematic biases, but were unable to identify any consistent issues. Overall, annotations for 1,015 TMA core images from 42 of the 47 involved annotators in the explanation annotation phase were included.

All annotations were performed using the online annotation tool PlainSight [49]. Further details on the data selection is available in **Supplementary Figure S.5**.

## Data Preparation

After the raters completed the explanatory annotations, explanations provided via the free text option had to be standardized. This was achieved by mapping them to their nearest equivalent in the ontology.

Sorted by the time of polygon creation, empty explanation fields were filled with the next available explanation. This reflected the expected behavior outlined in the instruction video provided to the raters: polygons were drawn first, afterwards the explanations were selected.

If multiple explanations were selected for a polygon, each explanation was included in the data as a copy of the original polygon.

For our model training, we created segmentation masks for each annotator and TMA core image pairs. As the explanations for each Gleason pattern were annotated separately on single grade images, we drew the annotations of each single grade image in the order they were created, with the single grade images themselves being sorted by their corresponding Gleason pattern in ascending order. Annotations that were provided later therefore override previous annotations of the same or lower Gleason patterns that share the same pixels.

The tree-structure of the ontology allowed us to create three datasets from our annotations (sub-explanations, explanations and Gleason patterns, see **Figure 8**), by remapping the sub-explanation annotations upwards in the ontology. This was deemed necessary for the model development, due to concerns of the sample size for the sub-explanations and the issue of class imbalance.

For both training and label analysis, we used merged grade images to review the complete grading and interpretations provided by the pathologists for the TMA core images.

## Model Development

To develop a pathologist-like, inherently explainable AI system (Gleason XAI) for detecting Gleason patterns on TMA cores, we employed a concept bottleneck strategy [30], predicting the explanations directly for each pixel, with the ability to later remap them to their corresponding Gleason pattern. This provides inherent explainability by basing the decision for a Gleason pattern solely on the predictions of the associated explanations, which can in turn be verified by expert pathologists in contrast to black box decisions on the Gleason pattern. Inspired by recent positive results [50], we selected a U-Net [31] with an Image-Net [51]-pretrained EfficientNet-B4 [52] encoder as our segmentation architecture.

We trained models on all three levels of our ontology (see **Figure 8**). At each level of the class hierarchy, an additional *benign tissue* class was included to account for unannotated regions.

As the white background of the slide would be falsely labeled as *benign tissue*, we used Otsu's thresholding [53] followed by morphological closing and opening operations to detect and mask out these regions during the loss-computation (see **Supplementary Figure S.6a**).

During inference the entire image was processed, however, the loss functions and the metrics were computed using only foreground pixels. For our segmentation visualizations, we also masked out the background pixels.

Due to high interobserver variability in the annotations, particularly for the explanations and sub-explanations, using majority voting or more sophisticated annotation merging approaches like STAPLE to obtain definite per-pixel labels was not feasible. Instead, we used a soft label approach, obtained by combining the annotations into a per-pixel annotation distribution (see **Supplementary Figure S.6b**). This method preserved human uncertainty in the labeling process and retained annotations for minority classes or cases where multiple explanations might be feasible.

When training segmentation models with majority-voted labels, often a combination of cross-entropy loss and variants of the Dice loss [54] is used. However, the Dice loss does not apply to soft labels. Let $p^{seg}, y^{seg} \in \mathbb{R}^{NxCxP}$ represent the predictive segmentation distribution and soft label for $N$ images, with $C$ classes, and $P$ foreground pixels flattened into a vector, normalized over the class dimension. Further, let $\epsilon$ represent a small smoothing constant.

We extended the Macro Dice loss to handle distributional inputs and targets as follows:

$$SoftDiceLoss(p^{seg}, y^{seg}) = 1 - \frac{1}{C} \sum_{c \in 1:C} \left( \frac{2 \sum_{p \in P, n \in N}(p^{seg}_{n,c,p} \cdot y^{seg}_{n,c,p}) + \epsilon}{\sum_{p \in P, n \in N}(p^{seg}_{n,c,p} + y^{seg}_{n,c,p}) + \epsilon} \right)$$

This loss is similar to the one proposed by *Wang et.al.* [42], however, we found that our loss function provided better and more consistent results in terms of our metrics, between different seedings.

For comparison, we also included models trained with the cross-entropy and class-averaged Dice loss[55] on majority-voted labels. These loss-functions were only computed over foreground pixels that possess an unambiguous majority vote, as there were cases where all three annotating pathologists disagreed on the class for a pixel.

Each model was trained three times with different seeds, using a batch size of 12 for 200 epochs. The starting learning rate was set to 5e-5, and it was reduced by a factor of three if the validation loss did not decrease for two consecutive epochs. The $L_2$ parameter regularization coefficient was set to 0.02. Both parameters were found through hyperparameter optimization using the *optuna*[56] library. We used AdamW [57] as optimizer with default parameters ($\beta_1 = 0.99, \beta_2 = 0.9$). Models were saved after each epoch, and the epoch with the lowest validation loss was selected for testing.

The TMA core images possessed resolutions between 2232x2215 px² and 5632x5632 px², stemming from different scanners and different physical sizes of the images. As the individual datasets had different pixel spacings (Gleason19 Challenge: 0.25 $\frac{\mu m}{px}$, Harvard Dataverse: 0.23 $\frac{\mu m}{px}$, TissueArray.com: 0.5455 $\frac{\mu m}{px}$), we tested multiple resolutions for segmentation performance and then bi-cubically interpolated all images to a common physical pixel side length of 1.392 µm/px. At this resolution the smallest images filled a 512x512 px² patch and the largest images reached 1358x1358 px². We augmented the images during training using the light augmentations without stain-normalizations, as recommended by *Tellez et al.* [58] and extracted random patches of size 512x512 px², while avoiding patches consisting only of background. For the validation set we always extracted the central 512x512px² patch of each image. At test time we utilized the computationally more expensive sliding window approach, implemented by *MONAI* [59], extracting 512x512px² patches, with 50% overlap and gaussian weighted averaging with default parameters.

The dataset was randomly split into training, validation and test datasets, comprising 70%, 15% and 15% of the TMA cores, respectively. We optimized the assignment of images to the training, validation and test split to minimize the L1-norm (see [Model Evaluation](#)) between the class distributions of the pixels between the splits. The class distributions are shown in **Supplementary Figure S.7.**

## Statistics

### Pathologists' caseload

The participating pathologists were requested to provide an estimate of the number of prostate cancer patients they examined per week, with their responses provided in ranges. To address the issue of overlapping ranges, a representative value was determined for each pathologist by calculating the mean of their submitted range. The median of these means was reported in the results.

## Interobserver Agreement

To analyze the annotator concordance while accounting for agreement occurring by chance, we calculated Fleiss' kappa [32] within each rater group and label using the statsmodels [60] implementation.

For this calculation, we binarized the pathologists' decisions by evaluating, for each image and label, whether the label was used by the annotator, resulting in nine decisions per image and annotator. Fleiss' kappa values were then calculated for each label within each group. To assess overall agreement within each group, we calculated Fleiss' kappa across all decisions of the group regardless of the labels. Additionally, we quantified the overall agreement for each label by calculating Fleiss' kappa across all decisions for that label in all TMA core images.

Furthermore, we also evaluated the interobserver agreement at the pixel level by counting the number of pixels assigned to each class and distributing them based on the number of agreeing annotators. However, since pixel-level data cannot be assumed to be independently distributed, we did not calculate Fleiss' kappa for this analysis.

## Model Evaluation

In our setting, which used soft labels for segmentation tasks, traditional metrics like Dice score, Intersection-over-Union, accuracy, and ROC curves offer an incomplete picture, as they rely on hard labels, and do not take into account the reviewers' clinical uncertainty.

Therefore, we reported the complement of the $mIoU^D(\alpha = 0.5, \beta = 0.5)$ loss by *Wang et al.* [42], a segmentation Dice loss that was developed for soft labels, as our primary metric (referred to as Macro SoftDice for brevity). Importantly, this loss averages over all classes, which is important for datasets containing large class imbalances.

Additionally, we focused on evaluating the predictive calibration by measuring the divergence between the predictive and target distributions. Contrary to hard label-based metrics, this approach accounted for pathologists' diagnostic uncertainty. We reported the $L_1$-norm normalized to the range (0,1), from now on called $L_1$-norm, between the predictive and target distributions:

$$L_1(p^{seg}, y^{seg}) = \frac{1}{2NCP} \sum_{n,c,p=1}^{N,C,P} |p_{n,c,p} - y_{n,c,p}|$$

A pathologist will most likely interact with visualization of the most probable class per pixel, as model explanations need to be simple and efficient in clinical practice [16]. Therefore, to maintain compatibility with adjacent literature and to provide intuitively understandable metrics, we also reported the Dice and the Macro Dice scores, computed globally over all pixels. These were calculated between the pixel-wise maximum of our predictive distribution and the majority vote of the annotation distribution as our secondary metrics. We also included a confusion matrix of the explanation predictions.

In our dataset, a unique majority vote could not be defined for pixels, where all three pathologists disagreed. Therefore, we computed these metrics only over foreground pixels with an unambiguous majority vote. The primary goal of this paper was to develop a pathologist-like, inherently explainable segmentation model for Gleason patterns on TMAs. Accordingly, we also evaluated the performance of the models in Gleason pattern segmentation. We compared models trained on the (sub-)explanations of our ontology to models trained solely on the Gleason patterns. For our models trained on the (sub-)explanations, we remapped the predictions to the Gleason patterns, by aggregating the corresponding probabilities of the predictions, summing up the predicted probabilities of the (sub-)explanations that — according to our predefined ontology — corresponded to a Gleason pattern.

We reported the mean and standard deviation of the metrics, averaged over three runs.

# Data Availability Statement

Explanation annotations based on all three datasets and TMA core images of the TissueArray.com dataset used in this study will be made publicly available upon publication. TMA core images of the Gleason 19 challenge are publicly available at https://gleason2019.grand-challenge.org/Register/. TMA core Images of the Arvaniti et al. Harvard Dataverse dataset are publicly available at https://doi.org/10.7910/DVN/OCYCMP.

# Code Availability Statement

The code for the model development and the statistical analyses will be made publicly available on GitHub upon publication.

# Authors Contributions Statement

SLP drafted the initial version of the manuscript. SLP conducted the recruitment and data collection phase. HM, GM and SH wrote the version sent out to the other authors of the manuscript. SH and TCB were involved in the supervision of the project. SLP, HM, TCB, GM, SH and TJB conceived the project. GM and HM conducted all analyses. TJB supervised the project from initiation to submission and takes responsibility for the published data as well as all analyses conducted. All authors provided clinical and/or machine learning expertise and contributed to the interpretation of the results and critically revised the manuscript.

# Literature

1.  Ferlay, J. *et al.* Cancer statistics for the year 2020: An overview. *Int. J. Cancer* (2021) doi:10.1002/ijc.33588.

2.  Gleason, D. F., Mellinger, G. T. & Veterans Administration Cooperative Urological Research Group. Prediction of Prognosis for Prostatic Adenocarcinoma by Combined Histological Grading and Clinical Staging. *J. Urol.* **197**, S134–S139 (1974).

3.  van Leenders, G. J. L. H. *et al.* The 2019 International Society of Urological Pathology (ISUP) Consensus Conference on Grading of Prostatic Carcinoma. *Am. J. Surg. Pathol.* **44**, e87 (2020).

4.  Epstein, J. I. *et al.* The 2019 Genitourinary Pathology Society (GUPS) White Paper on Contemporary Grading of Prostate Cancer. *Arch. Pathol. Lab. Med.* **145**, 461–493 (2021).

5.  Egevad, L., Granfors, T., Karlberg, L., Bergh, A. & Stattin, P. Prognostic value of the Gleason score in prostate cancer. *BJU Int.* **89**, 538–542 (2002).

6.  Epstein, J. I. *et al.* The 2014 International Society of Urological Pathology (ISUP) Consensus Conference on Gleason Grading of Prostatic Carcinoma: Definition of Grading Patterns and Proposal for a New Grading System. *Am. J. Surg. Pathol.* **40**, 244–252 (2016).

7.  Epstein, J. I. Prostate cancer grading: a decade after the 2005 modified system. *Mod. Pathol.* **31**, (2018).

8.  Burchardt, M. *et al.* Interobserver reproducibility of Gleason grading: evaluation using prostate cancer tissue microarrays. *J. Cancer Res. Clin. Oncol.* **134**, 1071–1078 (2008).

9.  Kartasalo, K. *et al.* Artificial Intelligence for Diagnosis and Gleason Grading of Prostate Cancer in Biopsies-Current Status and Next Steps. *Eur Urol Focus* **7**, 687–691 (2021).

10. Bulten, W. *et al.* Artificial intelligence assistance significantly improves Gleason grading of prostate biopsies by pathologists. *Mod. Pathol.* **34**, 660–671 (2021).

11. Bulten, W. *et al.* Artificial intelligence for diagnosis and Gleason grading of prostate

cancer: the PANDA challenge. *Nat. Med.* **28**, 154–163 (2022).

12. Marginean, F. *et al.* An Artificial Intelligence-based Support Tool for Automation and Standardisation of Gleason Grading in Prostate Biopsies. *Eur Urol Focus* **7**, 995–1001 (2021).

13. Nir, G. *et al.* Automatic grading of prostate cancer in digitized histopathology images: Learning from multiple experts. *Med. Image Anal.* **50**, 167–180 (2018).

14. Karimi, D. *et al.* Deep Learning-Based Gleason Grading of Prostate Cancer From Histopathology Images-Role of Multiscale Decision Aggregation and Data Augmentation. *IEEE J Biomed Health Inform* **24**, 1413–1426 (2020).

15. Hayashi, Y. Black Box Nature of Deep Learning for Digital Pathology: Beyond Quantitative to Qualitative Algorithmic Performances. *Artificial Intelligence and Machine Learning for Digital Pathology* 95–101 (2020).

16. Tonekaboni, S., Joshi, S., McCradden, M. D. & Goldenberg, A. What Clinicians Want: Contextualizing Explainable Machine Learning for Clinical End Use. (2019).

17. Goodman, B. & Flaxman, S. European Union regulations on algorithmic decision making and a 'right to explanation'. *AI Mag.* **38**, 50–57 (2017).

18. Understanding artificial intelligence ethics and safety: A guide for the responsible design and implementation of AI systems in the public sector. doi:10.5281/zenodo.3240529.

19. Selvaraju, R. R. *et al.* Grad-CAM: Visual Explanations from Deep Networks via Gradient-Based Localization. https://ieeexplore.ieee.org/document/8237336.

20. Bach, S. *et al.* On Pixel-Wise Explanations for Non-Linear Classifier Decisions by Layer-Wise Relevance Propagation. *PLoS One* **10**, e0130140 (2015).

21. Gunashekar, D. D. *et al.* Explainable AI for CNN-based prostate tumor segmentation in multi-parametric MRI correlated to whole mount histopathology. *Radiat. Oncol.* **17**, 1–10 (2022).

22. Arvaniti, E. *et al.* Automated Gleason grading of prostate cancer tissue microarrays via deep learning. *Sci. Rep.* **8**, 1–11 (2018).

23. Marco Tulio Ribeiro University of Washington, Seattle, WA, USA, Sameer Singh

University of Washington, Seattle, WA, USA & Carlos Guestrin University of Washington, Seattle, WA, USA. 'Why Should I Trust You?' https://dl.acm.org/doi/10.1145/2939672.2939778 doi:10.1145/2939672.2939778.

24. M, G., V, A., S, M.-M. & H, M. Concept attribution: Explaining CNN decisions to physicians. *Comput. Biol. Med.* **123**, 103865 (2020).

25. Sauter, D. *et al.* Validating Automatic Concept-Based Explanations for AI-Based Digital Histopathology. *Sensors* **22**, (2022).

26. Finding and removing Clever Hans: Using explanation methods to debug and improve deep models. *Inf. Fusion* **77**, 261–295 (2022).

27. Tomsett, R., Harborne, D., Chakraborty, S., Gurram, P. & Preece, A. Sanity Checks for Saliency Metrics. *AAAI* **34**, 6021–6029 (2020).

28. Adebayo, J. *et al.* Sanity Checks for Saliency Maps. *Adv. Neural Inf. Process. Syst.* **31**, (2018).

29. The explainability paradox: Challenges for xAI in digital pathology. *Future Gener. Comput. Syst.* **133**, 281–296 (2022).

30. Koh, P. W. *et al.* Concept Bottleneck Models. *arXiv [cs.LG]* (2020).

31. Ronneberger, O., Fischer, P. & Brox, T. U-Net: Convolutional Networks for Biomedical Image Segmentation. (2015) doi:10.48550/ARXIV.1505.04597.

32. Fleiss, J. L. Measuring nominal scale agreement among many raters. *Psychol. Bull.* **76**, 378–382 (1971).

33. Landis, J. R. & Koch, G. G. The measurement of observer agreement for categorical data. *Biometrics* **33**, 159–174 (1977).

34. Shah, R. B. *et al.* Diagnosis of Gleason pattern 5 prostate adenocarcinoma on core needle biopsy: an interobserver reproducibility study among urologic pathologists. *Am. J. Surg. Pathol.* **39**, 1242–1249 (2015).

35. Zhou, M. *et al.* Diagnosis of 'Poorly Formed Glands' Gleason Pattern 4 Prostatic Adenocarcinoma on Needle Biopsy: An Interobserver Reproducibility Study Among Urologic Pathologists With Recommendations. *Am. J. Surg. Pathol.* **39**, 1331–1339

(2015).

36. Shah, R. B. *et al.* Diagnosis of 'cribriform' prostatic adenocarcinoma: an interobserver reproducibility study among urologic pathologists with recommendations. *Am. J. Cancer Res.* **11**, 3990–4001 (2021).

37. Reinke, A. *et al.* Understanding metric-related pitfalls in image analysis validation. (2023) doi:10.1038/s41592-023-02150-0.

38. Lipton, Z. C. The Mythos of Model Interpretability. (2016) doi:10.48550/ARXIV.1606.03490.

39. Shwartz-Ziv, R., Goldblum, M., Li, Y. L., Bruss, C. B. & Wilson, A. G. Simplifying Neural Network Training Under Class Imbalance. (2023).

40. Hansum, T. *et al.* Comedonecrosis Gleason pattern 5 is associated with worse clinical outcome in operated prostate cancer patients. *Mod. Pathol.* **34**, 2064–2070 (2021).

41. Chen, N. & Zhou, Q. The evolving Gleason grading system. *Chin. J. Cancer Res.* **28**, 58–64 (2016).

42. Wang, Z., Popordanoska, T., Bertels, J., Lemmens, R. & Blaschko, M. B. Dice Semimetric Losses: Optimizing the Dice Score with Soft Labels. in *Medical Image Computing and Computer Assisted Intervention – MICCAI 2023* 475–485 (Springer Nature Switzerland, 2023).

43. Angelopoulos, A. N. & Bates, S. A Gentle Introduction to Conformal Prediction and Distribution-Free Uncertainty Quantification. (2021).

44. Website. Iczkowski KA. Gleason grading. PathologyOutlines.com website. https://www.pathologyoutlines.com/topic/prostategrading.html. Accessed November 20th, 2023.

45. TissueArray.Com. *TissueArray.Com* http://www.tissuearray.com/FAQs.

46. Zhong, Q. *et al.* A curated collection of tissue microarray images and clinical outcome data of prostate cancer patients. *Sci Data* **4**, 170014 (2017).

47. Arvaniti, E. *et al.* Replication Data for: Automated Gleason grading of prostate cancer tissue microarrays via deep learning. Harvard Dataverse

https://doi.org/10.7910/DVN/OCYCMP (2018).

48. Warfield, S. K., Zou, K. H. & Wells, W. M. Simultaneous truth and performance level estimation (STAPLE): an algorithm for the validation of image segmentation. *IEEE Trans. Med. Imaging* **23**, 903–921 (2004).

49. Plainsight Technologies, Inc. *PlainSight*. (2023).

50. Isensee, F., Jaeger, P. F., Kohl, S. A. A., Petersen, J. & Maier-Hein, K. H. nnU-Net: a self-configuring method for deep learning-based biomedical image segmentation. *Nat. Methods* **18**, 203–211 (2021).

51. Deng, J. *et al.* ImageNet: A large-scale hierarchical image database. in *2009 IEEE Conference on Computer Vision and Pattern Recognition* 248–255 (2009).

52. Tan, M. & Le, Q. V. EfficientNet: Rethinking model scaling for convolutional Neural Networks. (2019) doi:10.48550/ARXIV.1905.11946.

53. Otsu, N. A Threshold Selection Method from Gray-Level Histograms. *IEEE Trans. Syst. Man Cybern.* **9**, 62–66 (1979).

54. Milletari, F., Navab, N. & Ahmadi, S.-A. V-Net: Fully Convolutional Neural Networks for Volumetric Medical Image Segmentation. in *2016 Fourth International Conference on 3D Vision (3DV)* 565–571 (IEEE, 2016).

55. Milletari, F., Navab, N. & Ahmadi, S.-A. V-Net: Fully Convolutional Neural Networks for Volumetric Medical Image Segmentation. (2016).

56. Akiba, T., Sano, S., Yanase, T., Ohta, T. & Koyama, M. Optuna. in *Proceedings of the 25th ACM SIGKDD International Conference on Knowledge Discovery & Data Mining* (ACM, New York, NY, USA, 2019). doi:10.1145/3292500.3330701.

57. Loshchilov, I. & Hutter, F. Decoupled Weight Decay Regularization. in *International Conference on Learning Representations* (2019).

58. Tellez, D. *et al.* Quantifying the effects of data augmentation and stain color normalization in convolutional neural networks for computational pathology. *Med. Image Anal.* **58**, 101544 (2019).

59. Cardoso, M. J. *et al.* MONAI: An open-source framework for deep learning in

healthcare. (2022).

60. Seabold, S. & Perktold, J. Statsmodels: Econometric and statistical modeling with python. in *Proceedings of the 9th Python in Science Conference* (SciPy, 2010). doi:10.25080/majora-92bf1922-011.

# Supplementary Material

## Initial Ontology

**Table S.1:** Explanatory Ontology translated back from the German Ontology, with the numbering used in result figures on the right.

| Gleason Pattern 3 | |
|---|---|
| • single, individual atypical glands separated from each other | - 3.01 |
| • atypical glands with an irregularly separated, ragged, poorly defined edge | - 3.02 |
| • atypical glands are looser than a nodule and are infiltrative | - 3.03 |
| • either minute or large and cyst-like atrophic atypical glands | - 3.04 |
| • atypical glands lying very closely together (with little stroma between adjacent atypical glands) | - 3.05 |
| • well-formed, relatively uniform atypical glands with evenly distributed lumina | - 3.06 |
| • compressed or angular atypical glands | - 3.07 |
| • atypical glands infiltrate between benign glands | - 3.08 |

| Gleason Pattern 4 | |
|---|---|
| • slit-like lumina | - 4.01 |
| • large atypical glands | - 4.02 |
| • irregular contours, jagged edges of atypical glands | - 4.03 |
| • atypical glands fused or grown together into cords or chains | - 4.04 |
| • irregular distribution of lumina | - 4.05 |
| • atypical glands very close together (with little or no stroma) | - 4.06 |
| • Cribriform | - 4.07 |
|     ○ larger than a normal prostate gland; tends to fragmentation | - 4.08 |
|     ○ confluent sheet of contiguous carcinoma cells with multiple glandular lumina that are easily visible at low power (objective magnification 10x) | - 4.09 |
|     ○ single or fused glandular structures connected to each other (no intervening stroma or mucin) | - 4.10 |
| • Hypernephroid pattern | - 4.11 |
|     ○ nests of clear cells resembling renal cell carcinoma | - 4.12 |
|     ○ small, hyperchromatic nuclei | - 4.13 |
|     ○ fusion of acini into more solid sheets with the appearance of back-to-back glands without intervening stroma | - 4.14 |
| • Glomeruloid pattern | - 4.15 |
|     ○ rare small cribriform variant resembling glomerulus structures of kidney | - 4.16 |
|     ○ contains a tuft of cells that is largely detached from its surrounding duct space except for a single point of attachment | - 4.17 |

| Gleason Pattern 5 | |
|---|---|
| • solid tumor cell clusters with nonpolar nuclei around a lumen | - 5.01 |
| • Presence of definite comedonecrosis (central necrosis) | - 5.02 |
|     ○ with intraluminal necrotic cells | - 5.03 |
|     ○ with karyorrhexis within papillary, cribriform spaces | - 5.04 |
| • Single cells | - 5.05 |
|     ○ forming cords | - 5.06 |
|     ○ with vacuoles (signet ring cells) lacking glandular lumina | - 5.07 |

# Annotator Agreement for Sub-Explanations

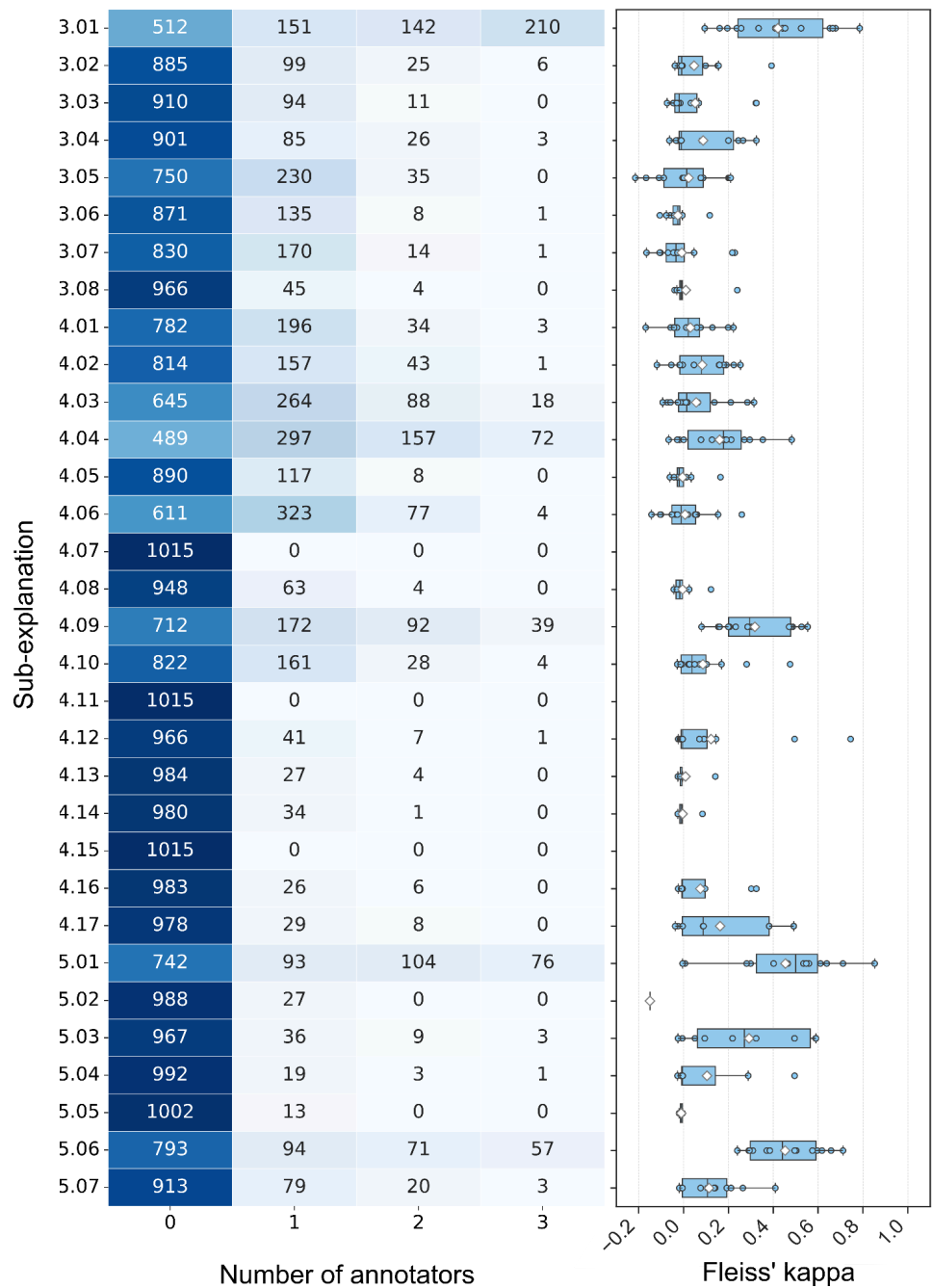

**Figure S.1:** Agreement of annotators for explanations on the image-level. Heatmap containing the number of TMA core images in which n out of the three annotators indicated the presence of the sub-explanations and the resulting Fleiss' kappa for groups of three raters (on the right). Dots represent the groups of annotators, diamonds the mean, boxes represent quartiles, and whiskers extend to the furthest datapoint within 1.5 of the inter-quartile range. As not all groups used all categories, the number of groups per category varies.

# Fleiss' Kappa Tables

**Table S.2:** Fleiss' kappa (κ) and 95% confidence interval (bootstrap: 10.000 resamples) within each label for Gleason patterns and explanations.

| Label | κ | 95% CI |
|---|---|---|
| Gleason Pattern | | |
| 3 | 0.784 | [0.751, 0.813] |
| 4 | 0.684 | [0.647, 0.719] |
| 5 | 0.786 | [0.749, 0.819] |
| Explanations | | |
| 3 - compressed glands | 0.145 | [0.096, 0.203] |
| 3 - individual glands | 0.710 | [0.676, 0.744] |
| 4 - cribriform glands | 0.431 | [0.384, 0.479] |
| 4 - glomeruloid glands | 0.329 | [0.203, 0.480] |
| 4 - poorly formed glands | 0.532 | [0.493, 0.571] |
| 5 - comedonecrosis | 0.347 | [0.229, 0.476] |
| 5 - cords | 0.532 | [0.475, 0.590] |
| 5 - groups of tumor cells | 0.549 | [0.500, 0.598] |
| 5 - single cells | 0.180 | [0.114, 0.270] |

**Table S.3:** Fleiss' kappa and 95% interval (bootstrap: 10.000 resamples) within each label for sub-explanations. If no value is given, the label didn't occur according to all raters.

| Sub-Explanation | κ | 95% CI | Sub-Explanation | κ | 95% CI |
| --- | --- | --- | --- | --- | --- |
| 3.01 | 0.577 | [0.535, 0.616] | 4.09 | 0.339 | [0.288, 0.397] |
| 3.02 | 0.214 | [0.146, 0.304] | 4.10 | 0.108 | [0.058, 0.173] |
| 3.03 | 0.059 | [0.016, 0.114] | 4.11 | - | [-, -] |
| 3.04 | 0.201 | [0.139, 0.284] | 4.12 | 0.156 | [0.070, 0.303] |
| 3.05 | 0.020 | [-0.11, 0.056] | 4.13 | 0.104 | [0.025, 0.221] |
| 3.06 | 0.022 | [-0.014, 0.089] | 4.14 | 0.016 | [-0.012, 0.129] |
| 3.07 | 0.020 | [-0.015, 0.074] | 4.15 | - | [-, -] |
| 3.08 | 0.059 | [0.006, 0.146] | 4.16 | 0.147 | [0.061, 0.259] |
| 4.01 | 0.075 | [0.033, 0.126] | 4.17 | 0.165 | [0.080, 0.262] |
| 4.02 | 0.116 | [0.076, 0.163] | 5.01 | 0.549 | [0.500, 0.600] |
| 4.03 | 0.149 | [0.104, 0.200] | 5.02 | -0.009 | [-0.013, -0.006] |
| 4.04 | 0.246 | [0.202, 0.292] | 5.03 | 0.271 | [0.155, 0.431] |
| 4.05 | 0.017 | [-0.014, 0.065] | 5.04 | 0.207 | [0.063, 0.495] |
| 4.06 | 0.026 | [-0.008, 0.063] | 5.05 | -0.004 | [-0.007, -0.002] |
| 4.07 | - | [-, -] | 5.06 | 0.532 | [0.473, 0.589] |
| 4.08 | 0.034 | [-0.006, 0.102] | 5.07 | 0.193 | [0.123, 0.286] |

**Table S.4:** Fleiss' kappa (κ) and 95% confidence interval (bootstrap: 10.000 resamples) within each group of raters.

|       | Gleason Patterns |                  | Explanations |                  | Sub-Explanations |                  |
|-------|------------------|------------------|--------------|------------------|------------------|------------------|
| Group | κ                | 95% CI           | κ            | 95% CI           | κ                | 95% CI           |
| 1.1   | 0.532            | [0.443, 0.620]   | 0.357        | [0.294, 0.425]   | 0.220            | [0.170, 0.281]   |
| 1.2   | 0.837            | [0.796, 0.896]   | 0.698        | [0.632, 0.756]   | 0.295            | [0.241, 0.353]   |
| 1.3   | 0.695            | [0.605, 0.774]   | 0.461        | [0.396, 0.532]   | 0.249            | [0.197, 0.308]   |
| 2.1   | 0.729            | [0.652, 0.805]   | 0.496        | [0.436, 0.556]   | 0.355            | [0.310, 0.401]   |
| 2.2   | 0.854            | [0.794, 0.907]   | 0.681        | [0.621, 0.737]   | 0.362            | [0.313, 0.415]   |
| 2.3   | 0.829            | [0.763, 0.882]   | 0.577        | [0.522, 0.632]   | 0.388            | [0.338, 0.440]   |
| 3.1   | 0.720            | [0.651, 0.786]   | 0.522        | [0.463, 0.580]   | 0.125            | [0.094, 0.162]   |
| 3.2   | 0.779            | [0.703, 0.839]   | 0.646        | [0.581, 0.708]   | 0.343            | [0.283, 0.408]   |
| 3.3   | 0.779            | [0.703, 0.837]   | 0.663        | [0.599, 0.722]   | 0.455            | [0.394, 0.516]   |
| 3.4   | 0.756            | [0.686, 0.817]   | 0.598        | [0.538, 0.658]   | 0.374            | [0.323, 0.427]   |
| 3.5   | 0.764            | [0.697, 0.822]   | 0.663        | [0.601, 0.719]   | 0.419            | [0.361, 0.477]   |
| 3.6   | 0.681            | [0.606, 0.751]   | 0.519        | [0.457, 0.581]   | 0.318            | [0.265, 0.374]   |
| 19.1  | 0.919            | [0.863, 0.954]   | 0.638        | [0.580, 0.696]   | 0.228            | [0.185, 0.276]   |
| 19.2  | 0.935            | [0.888, 0.965]   | 0.855        | [0.807, 0.895]   | 0.388            | [0.337, 0.443]   |

**Table S.5:** Fleiss' kappa per group and label for Gleason patterns and explanations. Missing values (NA) are due to explanations not used by the annotators in the group's data set.

| Label | Group | | | | | | | | | | | | | |
|---|---|---|---|---|---|---|---|---|---|---|---|---|---|---|
| | 1.1 | 1.2 | 1.3 | 2.1 | 2.2 | 2.3 | 3.1 | 3.2 | 3.3 | 3.4 | 3.5 | 3.6 | 19.1 | 19.2 |
| **Gleason Pattern** | | | | | | | | | | | | | | |
| 3 | 0.273 | 0.799 | 0.592 | 0.735 | 0.800 | 0.397 | 0.664 | 0.729 | 0.843 | 0.582 | 0.807 | 0.826 | 0.920 | 0.854 |
| 4 | 0.225 | 0.666 | 0.516 | 0.633 | 0.801 | 0.711 | 0.560 | 0.763 | 0.699 | 0.697 | 0.578 | 0.557 | 0.759 | 0.937 |
| 5 | 0.457 | 0.861 | 0.689 | 0.676 | 0.788 | 0.909 | 0.845 | 0.804 | 0.727 | 0.886 | 0.636 | 0.644 | 1.00 | 1.00 |
| **Explanation** | | | | | | | | | | | | | | |
| 3 - compressed glands | -0.005 | 0.091 | -0.043 | 0.277 | 0.100 | -0.009 | 0.102 | -0.006 | -0.103 | -0.109 | -0.130 | 0.210 | 0.399 | -0.036 |
| 3 - individual glands | 0.237 | 0.755 | 0.460 | 0.735 | 0.721 | 0.288 | 0.549 | 0.652 | 0.824 | 0.582 | 0.807 | 0.664 | 0.418 | 0.854 |
| 4 - cribriform glands | 0.400 | 0.611 | 0.347 | 0.536 | 0.244 | 0.559 | 0.213 | 0.422 | 0.606 | 0.243 | 0.404 | 0.261 | 0.280 | 0.463 |
| 4 - glomeruloid glands | -0.005 | -0.011 | 0.531 | 0.067 | -0.031 | 0.237 | 0.198 | -0.005 | -0.005 | NA | -0.008 | 0.796 | 0.457 | 0.658 |
| 4 - poorly formed glands | -0.010 | 0.412 | 0.201 | 0.528 | 0.699 | 0.301 | 0.486 | 0.594 | 0.661 | 0.624 | 0.519 | 0.437 | 0.690 | 0.858 |
| 5 - comedonecrosis | -0.053 | 0.655 | -0.006 | 0.152 | 0.495 | 0.618 | 0.096 | 0.590 | 0.852 | NA | -0.004 | 0.076 | NA | NA |
| 5 - cords | 0.309 | 0.503 | 0.596 | 0.293 | 0.575 | 0.711 | 0.291 | 0.658 | 0.240 | 0.617 | 0.294 | 0.371 | 0.385 | 0.496 |
| 5 - groups of tumor cells | 0.299 | 0.464 | 0.006 | 0.281 | 0.535 | 0.458 | 0.610 | 0.711 | 0.559 | 0.402 | 0.547 | 0.638 | 0.852 | -0.004 |
| 5 - single cells | 0.264 | -0.006 | 0.193 | 0.091 | 0.069 | 0.212 | -0.016 | 0.136 | 0.304 | -0.012 | -0.019 | 0.060 | 0.235 | NA |

**Table S.6:** Fleiss' kappa per group and label for sub-explanations. Missing values (NA) are due to explanations not used by the annotators in the group's data set (part 1).

| Sub-Explanation | Group | | | | | | | | | | | | | |
|---|---|---|---|---|---|---|---|---|---|---|---|---|---|---|
| | 1.1 | 1.2 | 1.3 | 2.1 | 2.2 | 2.3 | 3.1 | 3.2 | 3.3 | 3.4 | 3.5 | 3.6 | 19.1 | 19.2 |
| 3.01 | 0.237 | 0.409 | 0.161 | 0.653 | 0.422 | 0.094 | 0.195 | 0.524 | 0.678 | 0.453 | 0.785 | 0.334 | 0.257 | 0.665 |
| 3.02 | -0.005 | 0.153 | -0.030 | 0.156 | -0.020 | -0.009 | 0.042 | -0.030 | -0.019 | -0.024 | 0.039 | 0.098 | 0.392 | -0.009 |
| 3.03 | NA | 0.322 | NA | NA | -0.042 | 0.325 | -0.074 | -0.015 | NA | -0.046 | -0.023 | -0.030 | 0.033 | 0.068 |
| 3.04 | NA | 0.200 | NA | -0.027 | 0.244 | NA | 0.081 | -0.010 | -0.034 | 0.265 | -0.063 | -0.013 | -0.010 | 0.324 |
| 3.05 | NA | 0.087 | 0.014 | 0.200 | 0.205 | -0.004 | -0.215 | 0.087 | -0.109 | 0.077 | -0.167 | 0.210 | -0.088 | 0.002 |
| 3.06 | NA | -0.060 | -0.006 | NA | -0.031 | 0.117 | -0.106 | -0.020 | -0.024 | -0.077 | -0.042 | -0.005 | -0.026 | -0.027 |
| 3.07 | NA | -0.041 | -0.012 | 0.229 | -0.069 | NA | 0.047 | -0.046 | -0.103 | -0.165 | -0.107 | 0.217 | -0.020 | -0.027 |
| 3.08 | NA | -0.006 | -0.006 | NA | -0.005 | NA | -0.040 | -0.030 | NA | 0.240 | -0.015 | -0.004 | -0.015 | -0.018 |
| 4.01 | -0.046 | -0.058 | 0.014 | 0.076 | 0.031 | 0.199 | 0.012 | -0.030 | 0.029 | 0.222 | -0.043 | 0.129 | -0.169 | 0.059 |
| 4.02 | -0.119 | -0.055 | -0.018 | 0.190 | 0.047 | 0.158 | -0.004 | 0.179 | NA | -0.016 | 0.081 | 0.244 | 0.161 | 0.254 |
| 4.03 | -0.073 | 0.137 | 0.017 | 0.284 | -0.093 | -0.002 | 0.017 | 0.010 | -0.014 | 0.314 | 0.064 | -0.058 | -0.026 | 0.211 |
| 4.04 | -0.019 | 0.182 | -0.066 | 0.173 | 0.353 | -0.029 | 0.077 | 0.127 | 0.482 | 0.213 | 0.191 | 0.271 | 0.295 | 0.001 |
| 4.05 | -0.020 | -0.029 | -0.012 | -0.061 | 0.013 | 0.034 | -0.042 | -0.020 | -0.019 | -0.030 | -0.043 | 0.165 | -0.005 | 0.001 |
| 4.06 | -0.101 | -0.052 | -0.104 | -0.143 | 0.154 | 0.034 | 0-019 | -0.027 | 0.051 | 0.060 | 0.260 | 0.007 | 0.054 | -0.052 |
| 4.07 | NA | NA | NA | NA | NA | NA | NA | NA | NA | NA | NA | NA | NA | NA |
| 4.08 | -0.043 | -0.035 | -0.024 | -0.044 | -0.005 | 0.122 | -0.004 | -0.030 | NA | NA | NA | -0.009 | 0.025 | NA |

Table S.7: Fleiss' kappa per group and label for sub-explanations. Missing values (NA) are due to explanations not used by the annotators in the group's data set (part 2).

| Sub-Explanation | Group | | | | | | | | | | | | | |
|---|---|---|---|---|---|---|---|---|---|---|---|---|---|---|
| | 1.1 | 1.2 | 1.3 | 2.1 | 2.2 | 2.3 | 3.1 | 3.2 | 3.3 | 3.4 | 3.5 | 3.6 | 19.1 | 19.2 |
| 4.09 | 0.232 | 0.526 | 0.487 | 0.553 | 0.205 | 0.305 | 0.155 | 0.311 | 0.480 | 0.080 | 0.286 | 0.199 | 0.161 | 0.471 |
| 4.10 | -0.025 | 0.091 | 0.103 | 0.281 | -0.028 | -0.011 | 0.022 | 0.027 | 0.072 | 0.475 | 0.169 | 0.050 | -0.007 | -0.013 |
| 4.11 | NA | NA | NA | NA | NA | NA | NA | NA | NA | NA | NA | NA | NA | NA |
| 4.12 | -0.005 | NA | -0.024 | 0.091 | 0.745 | 0.145 | -0.016 | 0.071 | 0.495 | -0.004 | NA | -0.013 | -0.010 | -0.004 |
| 4.13 | -0.010 | NA | NA | 0.142 | -0.026 | -0.013 | -0.004 | -0.015 | NA | NA | NA | NA | -0.015 | NA |
| 4.14 | NA | 0.084 | -0.018 | -0.005 | -0.005 | -0.013 | -0.016 | -0.010 | NA | -0.004 | -0.008 | -0.026 | -0.026 | NA |
| 4.15 | NA | NA | NA | NA | NA | NA | NA | NA | NA | NA | NA | NA | NA | NA |
| 4.16 | -0.005 | -0.006 | -0.024 | -0.011 | -0.005 | 0.325 | 0.096 | -0.005 | -0.005 | Na | -0.008 | -0.009 | 0.302 | 0.324 |
| 4.17 | NA | -0.006 | 0.382 | 0.087 | -0.026 | -0.004 | -0.038 | NA | NA | NA | NA | 0.491 | 0.089 | 0.491 |
| 5.01 | 0.299 | 0.464 | 0.007 | 0.281 | 0.535 | 0.458 | 0.610 | 0.711 | 0.559 | 0.402 | 0.547 | 0.638 | 0.852 | -0.004 |
| 5.02 | -0.150 | NA | NA | NA | NA | NA | NA | NA | NA | NA | NA | NA | NA | NA |
| 5.03 | 0.324 | 0.589 | -0.006 | 0.051 | 0.495 | 0.218 | -0.025 | 0.590 | 0.590 | NA | NA | 0.094 | NA | NA |
| 5.04 | NA | -0.006 | NA | -0.027 | NA | 0.288 | -0.012 | NA | 0.495 | NA | -0.004 | -0.004 | NA | NA |
| 5.05 | NA | NA | NA | -0.005 | -0.020 | NA | NA | NA | -0.014 | -0.004 | NA | -0.004 | -0.015 | NA |
| 5.06 | 0.309 | 0.503 | 0.596 | 0.293 | 0.575 | 0.711 | 0.291 | 0.658 | 0.240 | 0.617 | 0.294 | 0.371 | 0.385 | 0.496 |
| 5.07 | 0.264 | -0.006 | 0.193 | 0.106 | 0.141 | 0.212 | -0.016 | 0.136 | 0.409 | -0.012 | 0.019 | 0.076 | -0.005 | NA |

# Additional model results

## Ablation study: Tree Loss

As the ontology has a natural hierarchical structure, we experimented with using a hierarchical tree-structured loss function for our soft labels that optimized a loss function for all levels of our ontology simultaneously. The models predictions on the lower levels of the ontology (explanations or sub-explanations), were additionally remapped to higher levels, thereby jointly optimizing the (sub-)explanation and Gleason pattern segmentations.

Let $y \in \mathbb{R}^C$ be a label distribution, with $y_i$ belonging to the i'th explanation or sub-explanation. Let $\phi(i)$ be the index of the corresponding class in the next higher ontology level (e.g. Gleason patterns for the explanations) belonging to the i'th explanation.

Given a loss-function $L$, we minimize the following loss-function for the prediction vector $p^{seg}, y^{seg} \in \Delta(N)$, with $\Delta(N)$ being the N-dimensional probability simplex:

$$TreeLoss_L(p^{seg}, y^{seg}) = \lambda L(p^{seg}, y^{seg}) + (1 - \lambda) L(p^{map}, y^{map})$$

, where $p_k^{map} = \sum_{i, \phi(i) = k} p_i$ being the sum of (sub-)explanation probabilities that belong to k'th Gleason pattern or explanation and $y^{map}$ defined analogously. We experimented with different values for $\lambda$ and set it to $\lambda = 0.5$. When training on sub-explanations the remapping is performed twice, first to the explanations and then to the Gleason patterns.

As we did not notice significant improvements using this method, we did not include it into our main results, however some interesting observations can be found in **Supplementary Figure S.2**

## Results on the sub-explanations

As can be seen in **Supplementary Figure S.2**, when training on the sub-explanations, the quality of the segmentations dropped significantly, with the Macro Dice dropping below 0.11, while the Dice score remained at 0.67, even better than the best performing method for the explanations.

Our explanations for this behavior are the significant imbalance and inter-pathologist disagreements (see **Figure 2** and **Supplementary Figure S.1**) that were present in the sub-explanations.

Additionally, when evaluating the SoftDiceLoss and soft label cross-entropy loss on the Gleason patterns, a dramatic decrease in segmentation performance could be observed, which is unacceptable for clinical practice.

However, utilizing a hierarchical loss through our TreeLoss (see [Ablation study: Tree Loss](#)) that trains on all three levels of the ontology hierarchy at the same time, we are able to preserve much of the performance on the Gleason patterns. On the explanations and the sub-explanations, this approach performed worse than the other methods.

We suspect this behavior to stem from our loss formulation, which, when taking the gradient, gives a positive reinforcement to each explanation that constitutes to the correct label of the higher hierarchy level, therefore also enforcing wrong labels on the lower levels of the hierarchy, consequently leading to higher entropy predictions.

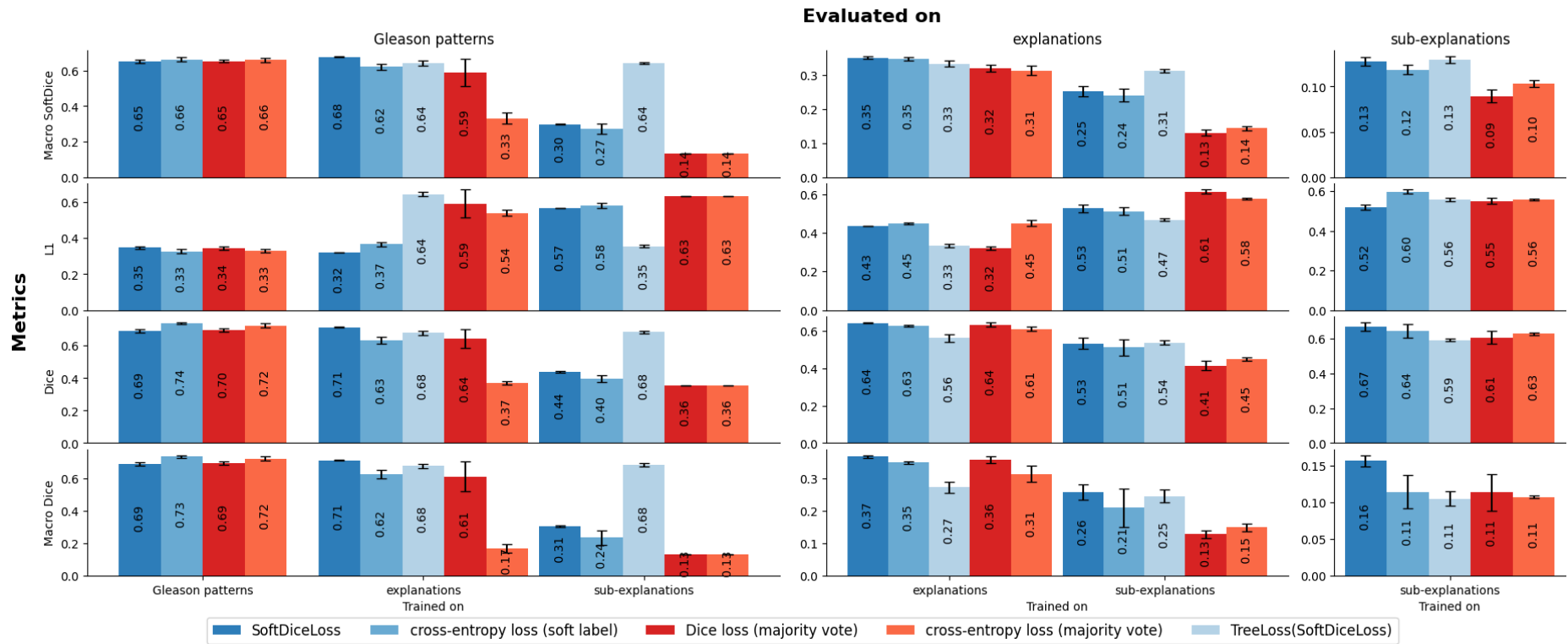

**Figure S.2**: Barplot of the results for models trained with different loss functions evaluated and trained on the Gleason patterns, the explanations as well as the original sub-explanations. Using our ontology we map the labels upwards in the ontology, comparing the performance of models trained at lower levels of the ontology with models directly trained on the Gleason patterns and explanations. We plot the mean and the standard deviation and repeat the mean inside of the bar plots.

# Numerical Results

**Table S.8**: Results for models trained with different loss functions evaluated and trained on the Gleason patterns, explanations and sub-explanations. Arrows indicate the direction of increasing performance. Each setting was trained 3 times, we report the mean and standard deviation for each metric. Abbr. 'patterns' : Gleason patterns, 'expl.': explanations

| Task | | | Metrics | | | | | | | |
|---|---|---|---|---|---|---|---|---|---|---|
| | | | soft label | | | | majority vote | | | |
| | | | Macro SoftDice ↑ | | $L_1$ ↓ | | Dice ↑ | | Macro Dice ↑ | |
| Evaluated | Trained | name | mean | std | mean | std | mean | std | mean | std |
| patterns | patterns | SoftDiceLoss | 0.651 | 0.011 | 0.346 | 0.011 | 0.691 | 0.010 | 0.688 | 0.011 |
| patterns | patterns | Dice loss (majority vote) | 0.653 | 0.010 | 0.344 | 0.011 | 0.696 | 0.014 | 0.694 | 0.014 |
| patterns | patterns | cross-entropy loss (majority vote) | 0.661 | 0.014 | 0.330 | 0.013 | 0.725 | 0.016 | 0.722 | 0.017 |
| patterns | patterns | cross-entropy loss (soft label) | 0.663 | 0.015 | 0.327 | 0.015 | 0.736 | 0.008 | 0.733 | 0.008 |
| patterns | expl. | SoftDiceLoss | 0.677 | 0.002 | 0.320 | 0.002 | 0.713 | 0.003 | 0.711 | 0.004 |
| patterns | expl. | Dice loss (majority vote) | 0.591 | 0.095 | 0.591 | 0.095 | 0.643 | 0.069 | 0.610 | 0.112 |
| patterns | expl. | TreeLoss(SoftDiceLoss) | 0.642 | 0.015 | 0.642 | 0.015 | 0.678 | 0.015 | 0.675 | 0.017 |
| patterns | expl. | cross-entropy loss (majority vote) | 0.333 | 0.039 | 0.540 | 0.021 | 0.369 | 0.012 | 0.168 | 0.033 |
| patterns | expl. | cross-entropy loss (soft label) | 0.621 | 0.023 | 0.365 | 0.016 | 0.632 | 0.025 | 0.625 | 0.032 |
| patterns | sub-expl. | SoftDiceLoss | 0.298 | 0.004 | 0.565 | 0.002 | 0.438 | 0.005 | 0.305 | 0.006 |
| patterns | sub-expl. | Dice loss (majority vote) | 0.135 | 0.000 | 0.630 | 0.000 | 0.356 | 0.000 | 0.131 | 0.000 |
| patterns | sub-expl. | TreeLoss(SoftDiceLoss) | 0.642 | 0.009 | 0.354 | 0.007 | 0.685 | 0.010 | 0.681 | 0.012 |
| patterns | sub-expl. | cross-entropy loss (majority vote) | 0.135 | 0.000 | 0.630 | 0.000 | 0.356 | 0.000 | 0.131 | 0.000 |
| patterns | sub-expl. | cross-entropy loss (soft label) | 0.273 | 0.036 | 0.580 | 0.018 | 0.397 | 0.028 | 0.235 | 0.056 |
| expl. | expl. | SoftDiceLoss | 0.351 | 0.004 | 0.433 | 0.001 | 0.643 | 0.004 | 0.367 | 0.005 |
| expl. | expl. | Dice loss (majority vote) | 0.319 | 0.012 | 0.319 | 0.012 | 0.635 | 0.013 | 0.358 | 0.013 |
| expl. | expl. | TreeLoss(SoftDiceLoss) | 0.333 | 0.010 | 0.333 | 0.010 | 0.562 | 0.023 | 0.273 | 0.020 |
| expl. | expl. | cross-entropy loss (majority vote) | 0.313 | 0.017 | 0.449 | 0.019 | 0.613 | 0.016 | 0.314 | 0.031 |
| expl. | expl. | cross-entropy loss (soft label) | 0.348 | 0.007 | 0.448 | 0.005 | 0.630 | 0.005 | 0.348 | 0.006 |
| expl. | sub-expl. | SoftDiceLoss | 0.252 | 0.019 | 0.526 | 0.024 | 0.534 | 0.037 | 0.257 | 0.029 |
| expl. | sub-expl. | Dice Loss (majority vote) | 0.131 | 0.011 | 0.613 | 0.013 | 0.414 | 0.031 | 0.129 | 0.015 |
| expl. | sub-expl. | TreeLoss(SoftDiceLoss) | 0.313 | 0.007 | 0.467 | 0.006 | 0.538 | 0.013 | 0.246 | 0.023 |
| expl. | sub-expl. | cross-entropy loss (soft label) | 0.240 | 0.023 | 0.512 | 0.023 | 0.514 | 0.054 | 0.210 | 0.071 |
| expl. | sub-expl. | cross-entropy loss (majority vote) | 0.144 | 0.007 | 0.577 | 0.005 | 0.450 | 0.010 | 0.149 | 0.014 |
| sub-expl. | sub-expl. | SoftDiceLoss | 0.127 | 0.006 | 0.519 | 0.015 | 0.666 | 0.031 | 0.156 | 0.009 |
| sub-expl. | sub-expl. | TreeLoss(SoftDiceLoss) | 0.129 | 0.004 | 0.558 | 0.012 | 0.590 | 0.010 | 0.105 | 0.012 |
| sub-expl. | sub-expl. | cross-entropy loss (soft label) | 0.119 | 0.006 | 0.598 | 0.013 | 0.643 | 0.050 | 0.114 | 0.028 |
| sub-expl. | sub-expl. | Dice loss (majority vote) | 0.090 | 0.008 | 0.550 | 0.018 | 0.605 | 0.046 | 0.113 | 0.030 |
| sub-expl. | sub-expl. | cross-entropy loss (majority vote) | 0.103 | 0.005 | 0.558 | 0.005 | 0.627 | 0.009 | 0.107 | 0.002 |

## Calibration Metric Discussion

When trained on the explanations, both the SoftDiceLoss and the soft label cross-entropy loss models achieved better or equal calibration on the Gleason patterns (see **Figure 5**) compared to the majority-voted cross-entropy models. The Dice loss, on the other hand, provided the best calibration when trained and evaluated on the explanations.

Evaluated on the Gleason patterns, soft label approaches trained on the explanations consistently outperformed the majority-voted approaches in terms of calibration, and were equally well calibrated when trained on the Gleason Patterns directly.

We therefore observe an exception to the superior performance of the soft labels in the $L_1$-norm, where models trained on Dice loss surpass those using soft label approaches at explanation level. Since the $L_1$-norm is not calculated on a per-class basis, we hypothesize that the high-confidence predictions from models trained on hard labels better accommodate the large class imbalances, leading to improved performance. Consensus among pathologists was highest for the most common classes (see **Figure 4**), resulting in labels that were largely equivalent to hard labels for these pixels. Conversely, the more conservative predictions from soft label approaches were less effective in terms of calibration for these majority classes. This highlights the importance of using diverse and balanced metrics when evaluating model performance.

# Gleason Pattern Masks

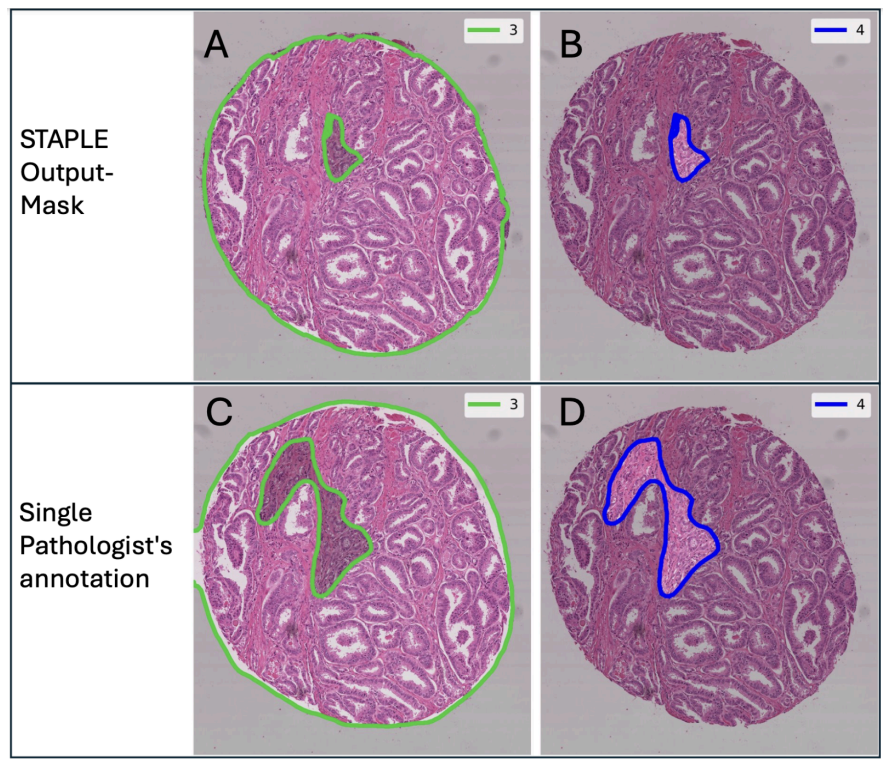

**Figure S.3: Review of STAPLE output masks**. In cases where the output masks did not cover biological meaningful patterns, the pathologists' annotation, which was closest to the output masks, was taken. Here a representative example is presented. Gleason pattern grade 3 areas are outlined in green and Gleason pattern grade 4 areas are outlined in blue. Corresponding areas are highlighted in a brighter color. STAPLE output mask with an inaccurate Gleason pattern grade 3 (A) and 4 (B) area. Single pathologist annotation with adjusted area for Gleason grade 3 (C ) and 4 (D).

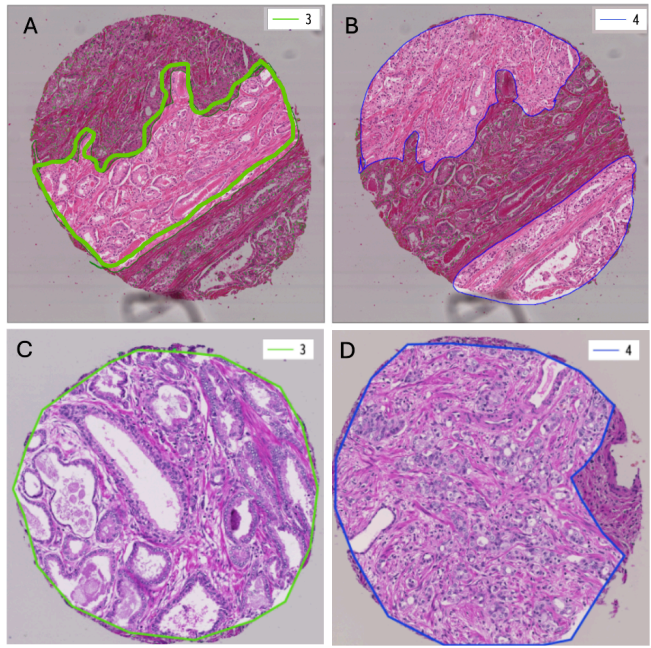

**Figure S.4: Representation of overall Gleason pattern grade annotation:** TMA core images with highlighted Gleason grade areas for a case of Gleason grade 4+3. Areas of A) Gleason grade 4 were highlighted and outlined in blue and for B) Gleason grade 3 were outlined in a separate image in green to avoid confusion. Were the Gleason grades the same, only a single area was highlighted i.e. in C) for Gleason 3+3 in green and in D) for 4+4 in blue.

## Data Selection

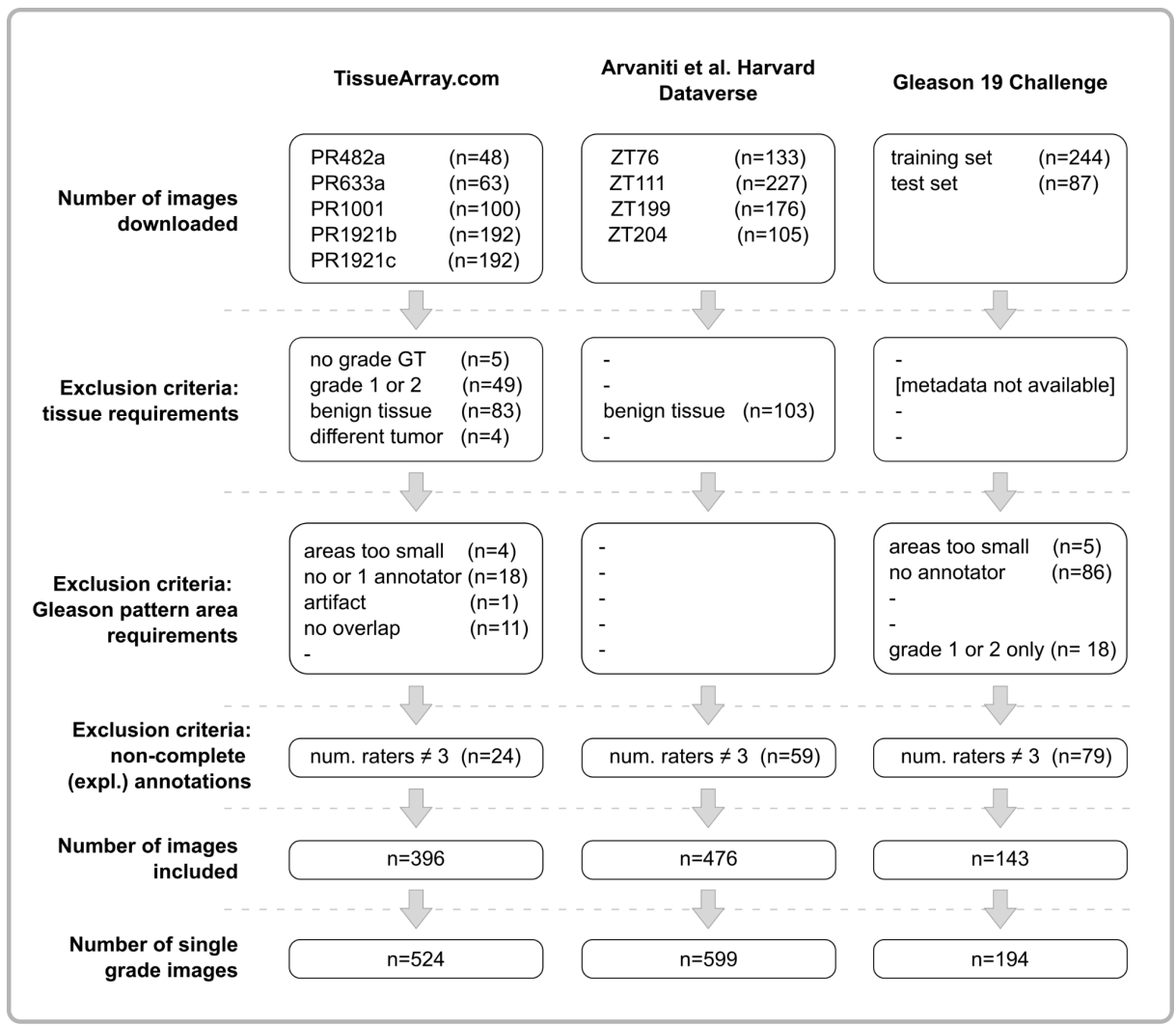

**Figure S.5** Slide selection process for the three datasets. '-': no images excluded by this filter, 'GT': ground truth.

# Label Generation

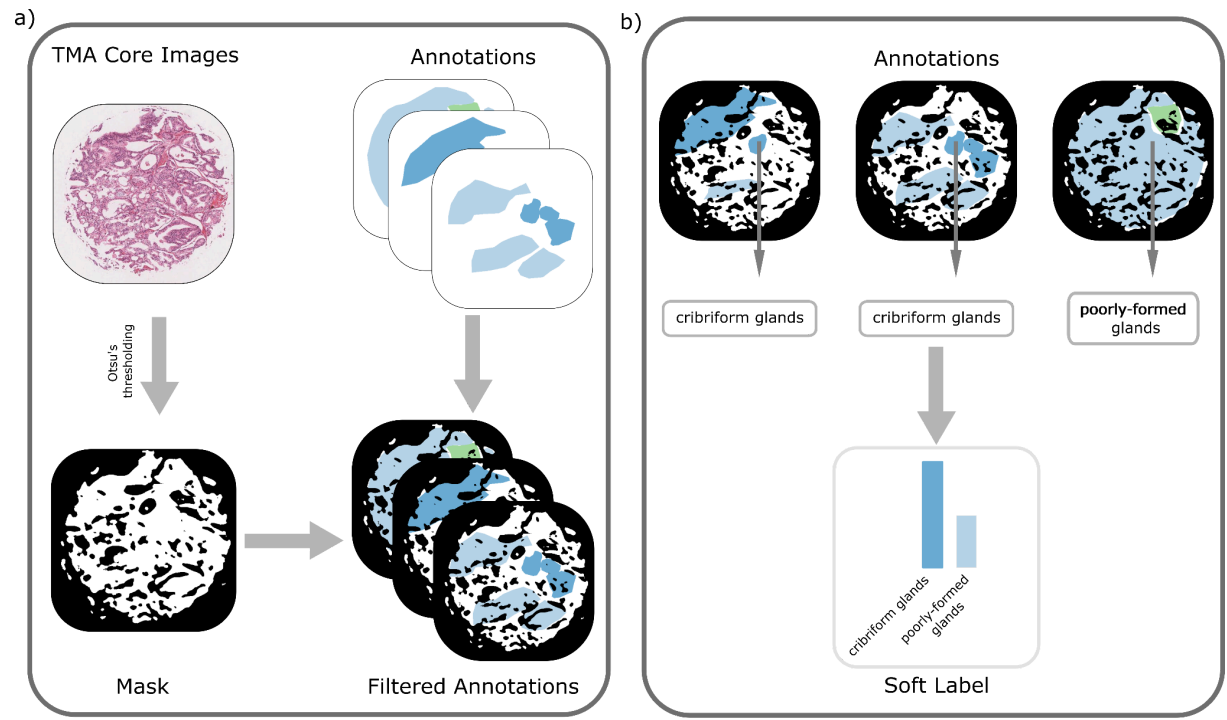

**Figure S.6** a) A foreground mask was extracted from the TMA core image by the use of Otsu's thresholding, followed by morphological operations. The annotations were filtered with the foreground mask to distinguish background pixels from benign tissue pixels. b) The per-pixel annotations of the annotators were combined into a (soft) per-pixel label-distribution. Here the distribution over the explanations for one pixel (indicated by the arrows) is shown.

# Data Split

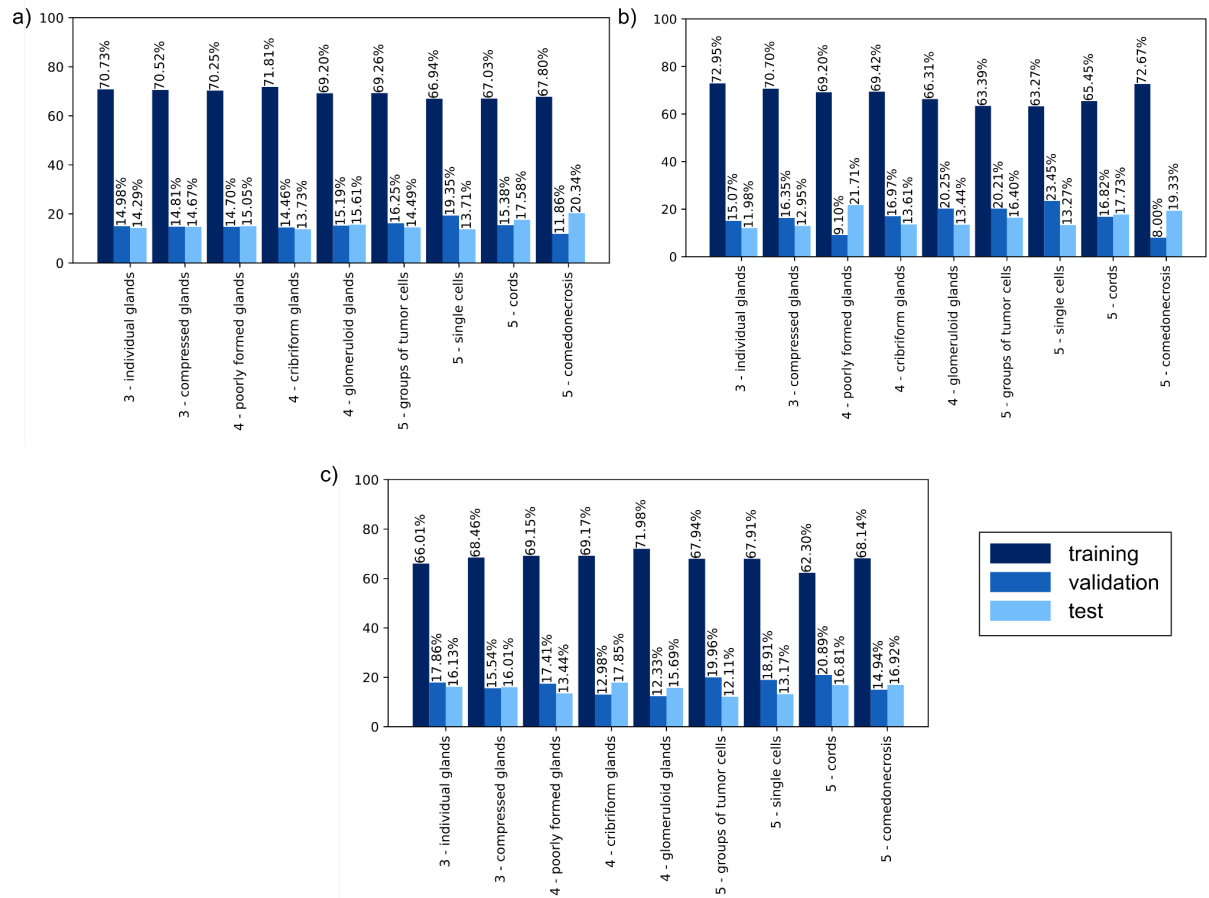

**Figure S.7:** Distribution of a) files, b) polygons, and c) pixels with explanations on the training, validation and test set.

# Ethics Declaration

Our research complies with all ethics regulations. The study's ethics vote was waived by a member of the ethics committee of the University Clinic Mannheim of the Medical Faculty of the University of Heidelberg, since our research involved no patients and we did not collect any patient data. Informed consent was obtained from all participating pathologists who performed the annotations of the publicly available data sets. We did not collect any data on sex and gender of the pathologists participating in our reader study. As compensation, we offered them the opportunity to be credited as a co-author of our work.

# Competing Interests

TJB owns a company that develops mobile apps (Smart Health Heidelberg GmbH, Heidelberg, Germany), outside of the scope of the submitted work. JNK declares consulting services for Bioptimus, France; Owkin, France; DoMore Diagnostics, Norway; Panakeia, UK; AstraZeneca, UK; Scailyte, Switzerland; Mindpeak, Germany; and MultiplexDx, Slovakia. Furthermore, he holds shares in StratifAI GmbH, Germany, Synagen GmbH, Germany, and has received a research grant by GSK, and has received honoraria by AstraZeneca, Bayer, Daiichi Sankyo, Eisai, Janssen, MSD, BMS, Roche, Pfizer and Fresenius. No other conflicts of interest are declared by any of the authors.